\documentclass{article}
\usepackage[utf8]{inputenc}
\usepackage{authblk}
\usepackage{setspace}
\usepackage[margin=1.25in]{geometry}
\usepackage{graphicx}
\graphicspath{ {./figures/} }
\usepackage{subcaption}
\usepackage{amsmath}
\usepackage{lineno}
\usepackage{color,amsmath,amssymb,graphicx,latexsym}
\usepackage{multirow}
\usepackage{ulem}
\usepackage{amsmath}
\usepackage{bm}
\usepackage{float}
\usepackage{booktabs}
\usepackage{threeparttable}
\usepackage{CJK}
\usepackage[whole]{bxcjkjatype} 
\usepackage{CJKutf8}
\usepackage[colorlinks,linkcolor=blue, anchorcolor=blue,citecolor=blue]{hyperref}
\usepackage{ulem,soul}
\usepackage{xcolor}

\usepackage[style=nejm, 
citestyle=numeric-comp,
sorting=none]{biblatex}

\title{Universal supercritical thermodynamics for black holes}

\author[1,3,*]{Shoucheng Wang}
\author[2,3,*]{Xinyang Li}
\author[3,4,5,$\dag$]{Yuliang Jin}
\author[3,4,6,$\ddag$]{Li Li}

\affil[1]{School of Science, Hunan Institute of Technology, Hengyang 421002, China.}
\affil[2]{Peng Huanwu Collaborative Center for Research and Education, Beihang University, Beijing 100191, China.}
\affil[3]{Institute of Theoretical Physics, Chinese Academy of Sciences, Beijing 100190, China.}
\affil[4]{School of Physical Sciences, University of Chinese Academy of Sciences, Beijing 100049, China.}
\affil[5]{Wenzhou Institute, University of Chinese Academy of Sciences, Wenzhou 325000, China.}
\affil[6]{School of Fundamental Physics and Mathematical Sciences, Hangzhou Institute for Advanced Study, UCAS, Hangzhou 310024, China.}
\affil[*]{Contributed equally to this work}
\affil[$\dag$]{yuliangjin@mail.itp.ac.cn}
\affil[$\ddag$]{liliphy@itp.ac.cn}

\date{}

\begin{document}

\maketitle

\begin{abstract}
We investigate thermodynamic crossovers for black holes in the supercritical regime beyond the critical point, where small and large black holes become indistinguishable from the conventional viewpoint. We establish a refined supercritical phase diagram that comprehensively characterizes the phases of small, large, and indistinguishable black holes, delineated by two supercritical crossover lines. The universal scaling laws of these crossover lines are fully verified using the thermodynamics of RN-AdS black holes in both the standard framework and the extended thermodynamic phase space, where the cosmological constant is treated as pressure, as well as in four other black hole systems. Analogies with supercritical crossovers observed in liquid-gas and liquid-liquid phase transitions are discussed. This work can be extended to more complex black hole backgrounds and offers valuable insights into the fundamental nature of black hole thermodynamics.
\end{abstract}


\section{Introduction}
Black hole thermodynamics extends classical concepts to extreme gravitational systems, revealing rich phase structures and critical phenomena. The interplay between geometry, thermodynamics, and holography underscores deep connections between gravity and statistical physics, with critical exponents universal across diverse systems. Of particular interest is the discovery of the charged Reissner–Nordstr\"{o}m (RN) black hole in the anti-de Sitter (AdS) spacetime with a negative cosmological constant $\Lambda$. This black hole admits a first-order small-black-hole/large-black-hole (SBH/LBH) phase transition, which is in many respects analogous to the liquid–gas phase transition { (LGPT)}~\cite{Chamblin:1999tk,Kubiznak:2012wp}. This analogy has triggered broad interest  and stimulated numerous studies on the critical behavior of black holes, greatly enriching the phase structure of black holes, \emph{e.g.} black hole chemistry~\cite{Kubiznak:2016qmn,Mann:2024sru,Karch:2015rpa} and QCD-like black hole phase~\cite{DeWolfe:2010he,Cai:2022omk,Grefa:2022sav,Zhao:2023gur,Jokela:2024xgz,Chen:2024ckb}. In particular, by treating $\Lambda$ and its conjugate quantity as thermodynamic variables associated with the pressure and volume, the RN-AdS black hole in this extended phase space displays classical critical behavior and is superficially analogous to the Van der Waals LGPT. In contrast, in the non-extended phase space with $\Lambda$  a fixed model parameter, the above analogy has not been well established, mainly due to the absence of good identification of pressure~\cite{Kubiznak:2012wp}.

Although black hole critical phenomena have been extensively studied for over two decades since~\cite{Chamblin:1999tk}, little progress has been made in exploring the properties of black holes above the critical point, and the thermodynamic nature of supercritical black holes (SCBHs) remains unknown. The behavior in the supercritical region is fundamentally different from that below the critical point. For the latter, thermodynamic response functions change discontinuously when crossing the coexistence line. For SCBHs, the thermodynamic crossover curve (Widom line) was constructed via the Ruppeiner geometry~\cite{Sahay:2017hlq,DasBairagya:2019nyv}, Lee-Yang zeros~\cite{Xu:2025jrk}, kinetic crossover~\cite{Li:2025tdd}, and maximal scaled variance~\cite{Zhao:2025ecg}. On the other hand, the supercritical dynamic crossover curve (Frenkel line) has been identified by transitions between distinct quasi-normal modes~\cite{Zhao:2025ecg}. However, these supercritical crossover lines do not exhibit universal behavior.

In this work, we investigate two recently proposed thermodynamic crossover lines, \(L^{\pm}\)~\cite{li2024thermodynamic}, which serve as boundaries of the SCBH region in the phase diagram. The $L^+$ and $L^-$ lines represent the loci where the thermodynamic response starts to deviate from LBH-like and SBH-like behavior, respectively. The region between the two lines is where the system no longer clearly resembles either phase, justifying the term ``indistinguishable''. Our primary goal is to extend the $L^\pm$ to black hole thermodynamics, which has lacked a systematic supercritical description. Unlike existing supercritical diagnostics (e.g., Widom or Frenkel lines) that do not show universal behaviour in black holes, the $L^\pm$ lines yield robust universal scaling laws, making them a natural tool for this purpose. Beyond universality verification, we envisage the scaling laws as a probe for beyond mean‑field criticality in future studies.

We establish universal scaling laws for the \(L^{\pm}\) lines near seven critical points in the following six black hole systems: (i) 4D RN-AdS black holes in the extended phase space, (ii) 4D RN-AdS black holes in the non-extended phase space, (iii) hairy black holes (with two critical points), (iv) RN-AdS black holes in higher dimensions, (v) 5D charged Gauss-Bonnet black holes, and (vi) Barrow black holes with fractal structure. We also discuss their analogies with LGPTs and liquid-liquid phase transitions (LLPTs).

\section{Results}
\subsection*{Comparison of various supercritical crossover lines for charged black holes in the extended phase space}

\begin{figure*}[hpt]
    \centering
     \includegraphics[width=1\columnwidth]{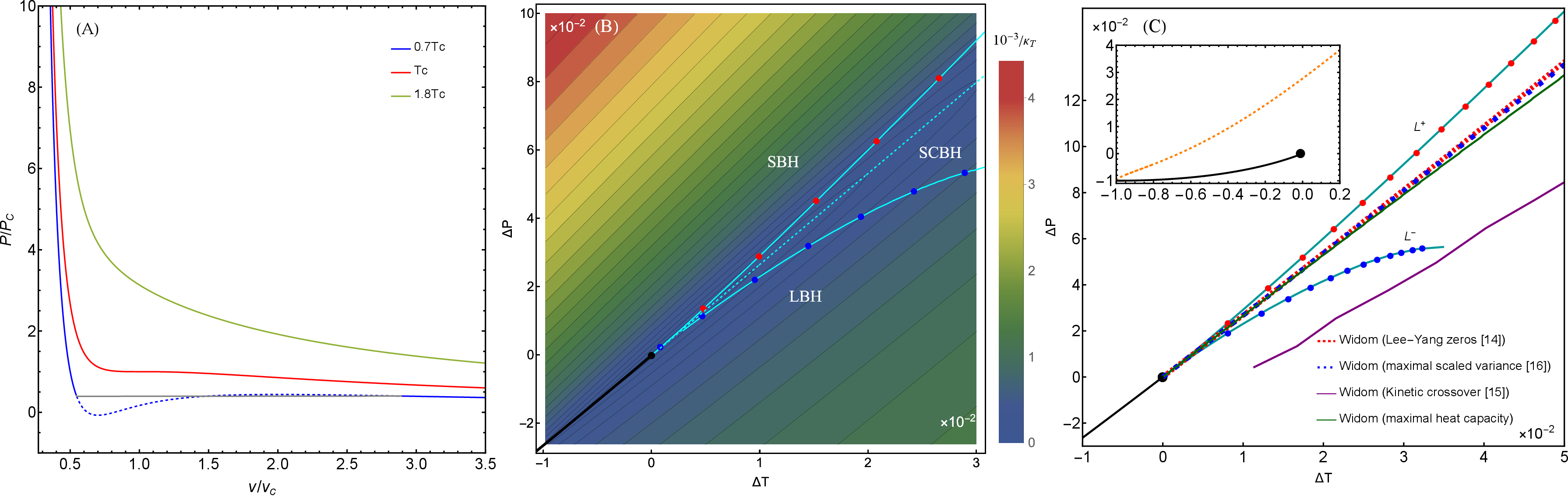}
    \caption{Supercritical thermodynamics for 4D RN-AdS black holes in  the extended phase space. (A) Typical EOSs in three phases. Beyond the
critical point $(T_c, v_c)$ is the supercritical regime. (B) Phase diagram in terms of rescaled axes, $\Delta P=P/P_c-1$ and $\Delta T=T/T_c-1$. The solid black and dashed cyan lines represent respectively the coexistence line and critical isochore. The solid black point marks the critical point.
The solid cyan line with red (blue) points represent the $L^+$ ($L^-$) line. The color map and contour lines are obtained via the isothermal compressibility. (C) Comparison with other supercritical crossover lines. The critical isochore is visually identical to the dashed Widom lines.
The inset presents the Frenkel line (dashed) determined via quasi-normal modes~\cite{Zhao:2025ecg}.  We have set $Q=1$.
}
    \label{fig:4DPV} 
\end{figure*}

We begin with the 4D RN-AdS black hole in the extended phase space.
It has arguably the simplest thermodynamic properties, since its equation of state (EOS. remarkably coincides with the Van der Waals EOS.
The EOS is given as~\cite{Kubiznak:2012wp},
\begin{equation}
    P(v, T)=\frac{T}{v}-\frac{1}{2\pi v^2}+\frac{2Q^2}{\pi v^4}\,,
    \label{eq:BHEPS_EOS}
\end{equation}
where  $P\equiv-\frac{1}{8\pi}\Lambda=\frac{3}{8\pi}\frac{1}{l^2}$ is the thermodynamic pressure (with $l$ the AdS radius)~\cite{Kubiznak:2012wp},  $v$ the specific volume and $Q$ the charge. This black hole undergoes a small (liquid-like) to large (gas-like) black hole phase transition (see Fig.~\ref{fig:4DPV}A). The critical point is located at, $T_c=\frac{\sqrt{6}}{18\pi Q}$, $v_c=2\sqrt{6}Q$ and $P_c=\frac{1}{96\pi Q^2}$. Note that the critical compressibility factor $Z_c = P_c v_c/T_c=3/8$ in this black hole is identical to $Z_c = 3/8$ in the Van der
Waals fluid. The inverse specific volume, $\rho = 1/v$, has been  identified as the number density of black hole molecules to measure the microscopic degrees of freedom~\cite{Wei:2015iwa}. The EOS Eq.~(\ref{eq:BHEPS_EOS}) exhibits standard critical scalings 
near the critical point. For example, the isothermal compressibility scales as, 
$\kappa_T\equiv-\frac{1}{v}\frac{\partial v}{\partial P}|_{T,Q}\propto |T-T_c|^{-\gamma}$, with  $\gamma=1$, along the critical isochore ($v=v_c$). The order parameter $\rho_g-\rho_l$ ($\rho_g$ is the density of the gas-like phase and $\rho_l$ the liquid-like phase) on the coexistence line behaves as $\rho_g-\rho_l\propto|T-T_c|^\beta$, with  $\beta=1/2$. 

Ref.~\cite{li2024thermodynamic} proposes a new definition for supercritical crossover lines, named $L^{\pm}$ lines, which account for the universal properties of critical points and the inherent symmetry of the phase diagram. These lines are operationally defined by locating the maxima of thermodynamic response functions along directions parallel to the critical isochore (the line of constant critical order parameter). Applying this definition to the  RN-AdS black hole yields the $L^{\pm}$ lines shown in Fig.~\ref{fig:4DPV}B. These lines segment the black hole's phase diagram into three characteristic zones: the SBH phase, a SCBH phase where SBH and LBH states are indistinguishable, and the LBH phase.

In addition to our work, alternative approaches have been proposed to analyze the supercritical properties of charged black holes, Eq.~\eqref{eq:BHEPS_EOS}, which are summarized in Fig.~\ref{fig:4DPV}C: 
the Widom line is a single crossover line defined by extrema of response functions~\cite{Xu:2025jrk,Li:2025tdd,Zhao:2025ecg}; the Frenkel line is also a single crossover line, which, however, does not originate from the critical point~\cite{Zhao:2025ecg}; Ruppeiner geometry provides an alternative geometric method to locate the Widom line~\cite{Wei:2015iwa}; Lee‑Yang zeros yield the Widom line via complex analytic continuation~\cite{Xu:2025jrk}; and kinetic crossover lines are dynamical in nature and often do not emanate from the critical point~\cite{Li:2025tdd}. In contrast to all these existing crossover lines, the $L^{\pm}$ construction provides a two‑line framework that divides the supercritical region into three regimes (SBH, indistinguishable, and LBH) and obeys universal scaling laws not shared by single‑line diagnostics. Its limitations include the need for a suitable order‑parameter (which is not obvious in the non-extended phase spaces), its thermodynamic (rather than dynamical) character, and weak dependence on the choice of the response function away from the critical point.

\begin{figure*}[hpt]
    \centering
    \includegraphics[width=0.44\columnwidth]{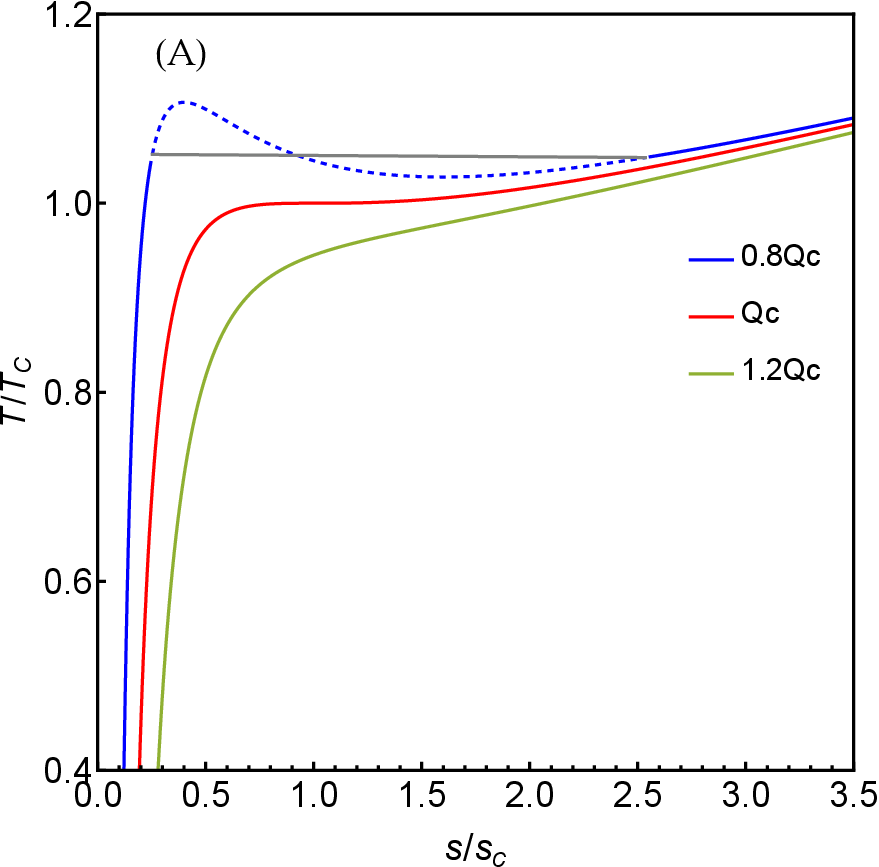}\qquad
    \includegraphics[width=0.51\columnwidth]{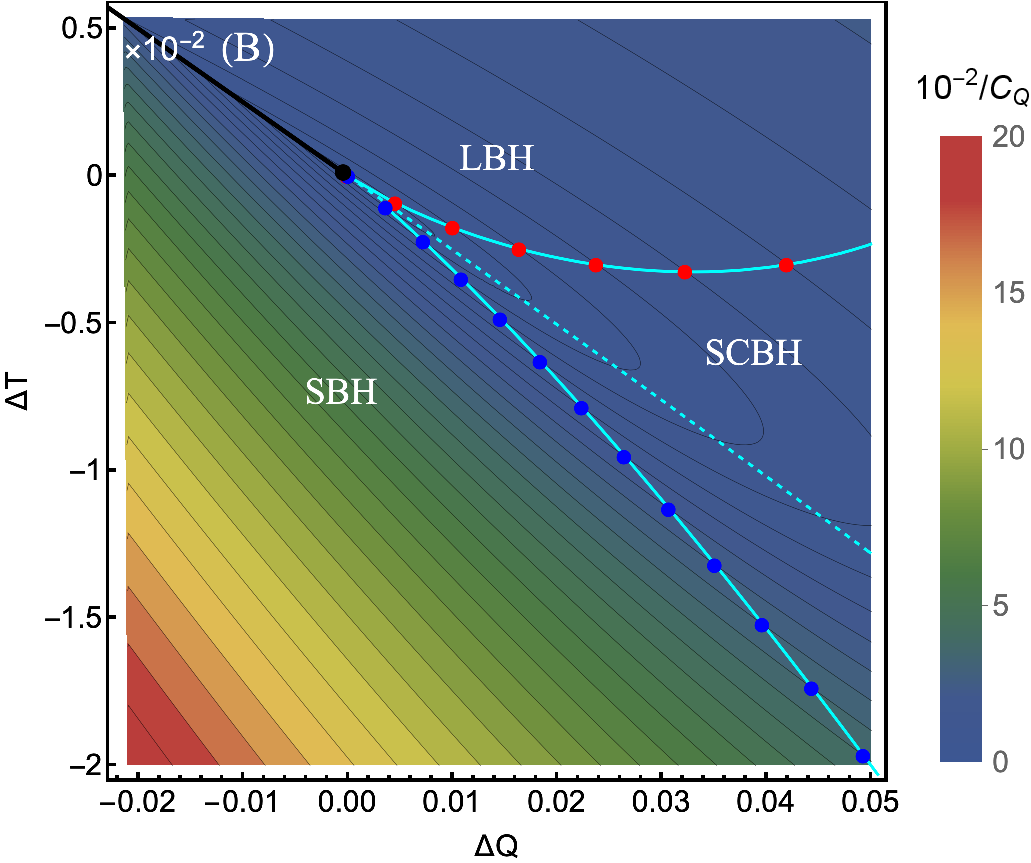}
    \caption{Supercritical thermodynamics for 4D RN-AdS black holes in  the non-extended phase space. (A) Typical EOSs in three phases. Beyond the
critical point $(T_c, Q_c)$ is the supercritical regime. (B) Phase diagram in terms of  rescaled $Q$ and $T$. The solid black and dashed cyan lines represent the coexistence line and the critical isentrope $S=S_c$, respectively; the latter serves as the supercritical extension of the coexistence line. The solid cyan line with red (blue) points marks the $L^+$ ($L^-$) line. The color map and contour lines are obtained according to the specific heat. We have set $l=\sqrt{3}$. 
    }
    \label{fig:4DTS} 
\end{figure*}

\subsection*{Universal scaling of $L\pm$ lines for seven critical points in charged black holes}

It has been suggested that $L^{\pm}$ lines should follow universal scaling laws~\cite{li2024thermodynamic}:
\begin{equation}
\delta P^{\pm} \propto (T-T_{\rm c})^{\beta + \gamma}\,,
\label{eq:H_scaling}
\end{equation}
and 
\begin{equation}
\delta \rho^{\pm} \propto (T-T_{\rm c})^{\beta}\,,
\label{eq:eta_scaling}
\end{equation}
near the critical temperature $T_{\rm c}$.
Here  $T$ is the control field,  $\rho$  the order parameter, $P$  the conjugated external field (ordering field), $\delta \rho^{\pm}$ ($\delta P^{\pm}$) measures the difference of $\rho$ ($P$) between the  $L^{\pm}$ lines and the critical isochore line ($\rho=\rho_{\rm c}$ line), and $\beta, \gamma$ are standard critical exponents whose values depend on the universality class. 
While these scaling relations have been examined in several non-gravitational systems~\cite{li2024thermodynamic,wang2024quantum,lv2024quantum}, it remains an open question whether gravitational effects preserve or alter this universality. A particularly crucial test would be applying these scalings to black hole systems that lack a clear Van der Waals fluid analogy, where the persistence of universal behavior is uncertain.

To address this question, we analyze the supercritical thermodynamics of seven distinct critical points in six black hole models (see supplementary materials (SM) Sec.~S1 for details). These models span a broad range of physical settings and phase structures, including cases with non-Van der Waals behavior. Remarkably, despite the introduction of non-trivial features such as quantum effects and fractal structures, the scaling relations defined in Eqs.~(\ref{eq:H_scaling}) and~(\ref{eq:eta_scaling}) are universally observed. This finding suggests that the proposed scalings possess a robustness that transcends both specific gravitational models and the Van der Waals analogy. The studied systems include the following black holes:

(i) Charged black holes in the extended phase space (RN-AdS$_4$), as detailed above. 

(ii) Charged black holes in the non-extended phase space (RN-AdS$_4$(NE)).
For such black holes,  where $\Lambda$ is a fixed model parameter, their correspondence to the liquid-gas system is debatable~\cite{Kubiznak:2012wp}. Nevertheless, a first-order SBH/LBH phase transition for the 4D RN-AdS black holes exhibits in the fixed charge ensemble (canonical ensemble)~\cite{Chamblin:1999tk}; see the $(T, Q)$ phase diagram in Fig.~\ref{fig:4DTS}A, where the first-order line is denoted by a solid black line with the critical point located at $(T_c=\frac{2}{\pi l \sqrt{6}}, \mu_c=\frac{1}{\sqrt{6}}, Q_c=\frac{l}{6})$. 

In this case, the LGPT concepts of thermodynamic pressure and its conjugate order parameter (density) are no longer applicable. Alternatively, Refs.~\cite{Chamblin:1999hg, Kubiznak:2012wp} propose to treat $Q$ as the control field and $T$ as the ordering field. Following this idea, we identify the black hole entropy $S$ (\emph{i.e.} Bekenstein-Hawking entropy) as the order parameter conjugated to $T$; {see also~\cite{HosseiniMansoori:2024bfi}}. Along the coexistence line, the entropy gap between the two phases is $s = S_2 - S_1$, which scales as $s \propto |Q-Q_{c}|^\beta$, with $\beta = 1/2$. Along the critical isentrope ($S = S_{c}$), the response function $C_Q \equiv T\frac{\partial S}{\partial T}|_Q $ follows the scaling $C_Q \propto |T-T_c|^{-\gamma}$ with $\gamma = 1$. Following a similar procedure as explained in Sec.~\ref{sec:definition}, the $L^\pm$ lines are determined (see Fig.~\ref{fig:4DTS}B).

\begin{figure}[hpt]
    \centering
    \includegraphics[width=0.48\columnwidth]{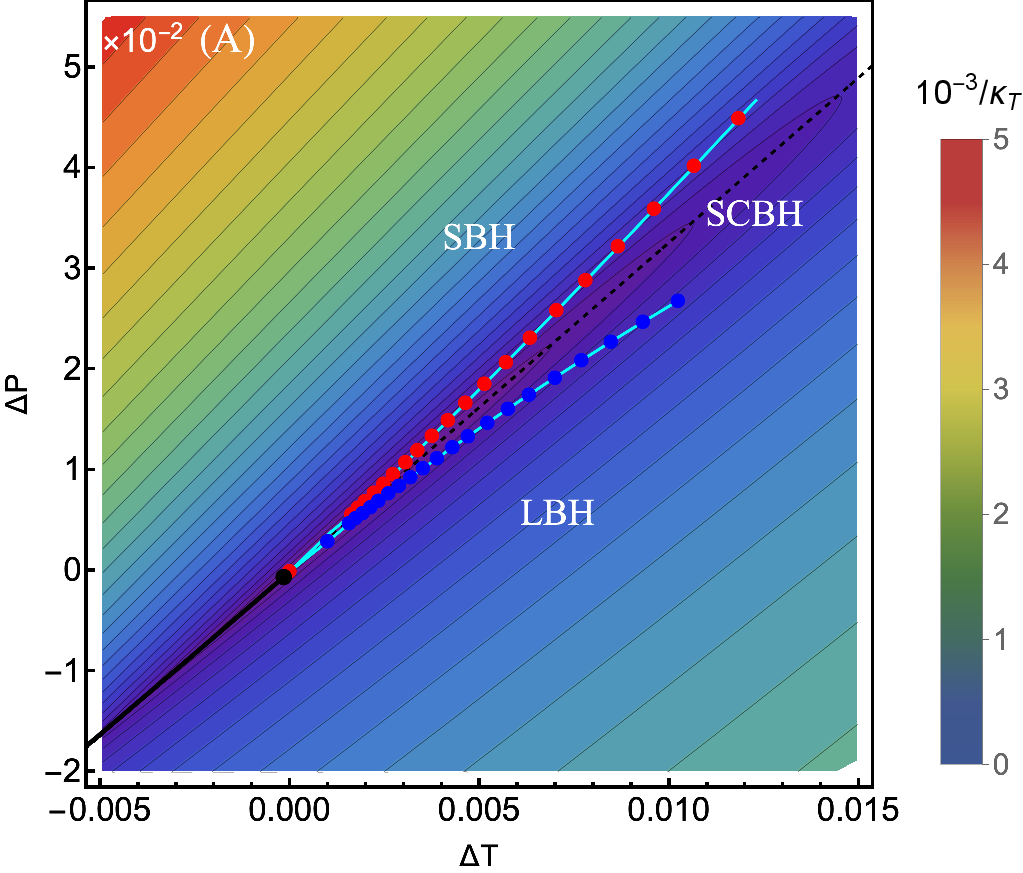}
    \includegraphics[width=0.48\columnwidth]{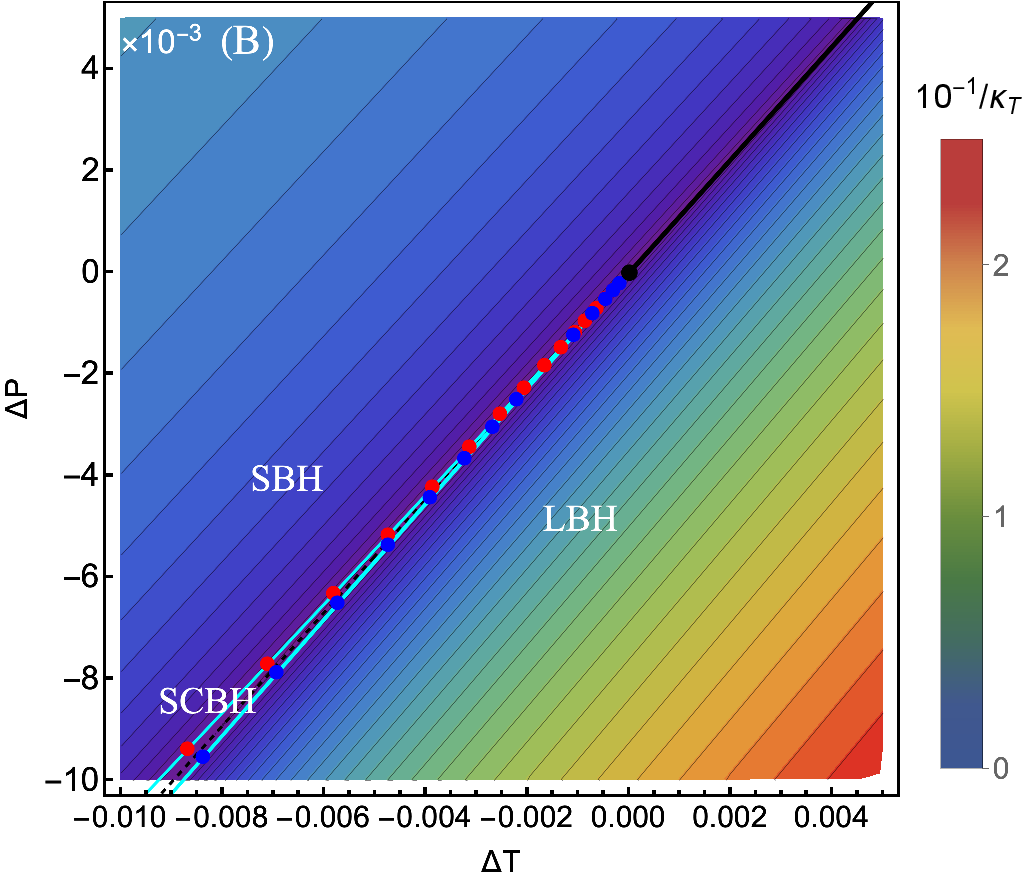}
    \caption{Supercritical thermodynamics for the hairy black hole~\cite{Astefanesei:2019ehu}, including (A) the RN-AdS–like critical point at large volume and low temperature, and (B) another critical point at small volume and high temperature. 
    }
    \label{fig:2CPs} 
\end{figure}

(iii) Hairy black holes with two critical points. The thermodynamics of a charged black hole with scalar hair in the extended phase space was studied in Ref.~\cite{Astefanesei:2019ehu}. For a fixed charge $Q$, the system exhibits two critical points: one resembling the RN-AdS type at large volume and low temperature, and another at small volume and high temperature. As $Q$ varies, the critical compressibility factor $Z_c=P_c v_c/T_c$ changes continuously, effectively tuning the nature of the dual fluid. This behavior clearly extends beyond the Van der Waals description, even though the critical exponents $\beta$ and $\gamma$ at each critical point retain mean-field values. Following the same procedure outlined earlier for the RN-AdS black hole, we numerically determine the supercritical crossover lines $L^{\pm}$, as shown in Fig.~\ref{fig:2CPs}.

(iv) RN-AdS black hole in higher dimensions.
The key thermodynamic features of the 4D RN-AdS black hole extend naturally to higher dimensions~\cite{Niu:2011tb}. We find that the characteristic phase behavior, analogous to that shown in Fig.~\ref{fig:4DPV}B, persists across dimensions.

(v) Charged Gauss-Bonnet black holes.
Higher-derivative curvature terms represent a common manifestation of quantum effects. Notably, the Gauss-Bonnet term arises naturally as a low-energy limit in string theory. Within the extended thermodynamic framework, we find that 5D charged Gauss-Bonnet black holes with spherical horizons~\cite{Cai:2013qga} indeed possess supercritical crossover lines 
$L^{\pm}$, analogous to those in Fig.~\ref{fig:4DPV}B.

(vi) Barrow black holes with fractal structures. The Barrow model introduces a fractal-inspired modification to black hole entropy, expressed as $S_{B}=S^{1+\delta/2}$~\cite{Barrow:2020tzx} which generalizes the Bekenstein-Hawking entropy $S$ to account for quantum gravitational spacetime foam. In this framework, the Barrow parameter $\delta$ ($0\le \delta\le 1$) measures the degree of quantum-gravitational deformation of the horizon area. Consistent with the first law, all other thermodynamic quantities are correspondingly rescaled by $\delta$. When applied to the 4D RN-AdS black hole, the Barrow-modified thermodynamics shows supercritical crossover lines qualitatively similar to those in Fig.~\ref{fig:4DPV}B. Notably, the conventional RN-AdS result is recovered in the limit $\delta=0$, illustrating how quantum-gravitational corrections ($\delta>0$) alter—yet do not fundamentally disrupt—the observed thermodynamic pattern.

\begin{figure}[hpt]
    \centering
    \includegraphics[width=0.48\columnwidth]{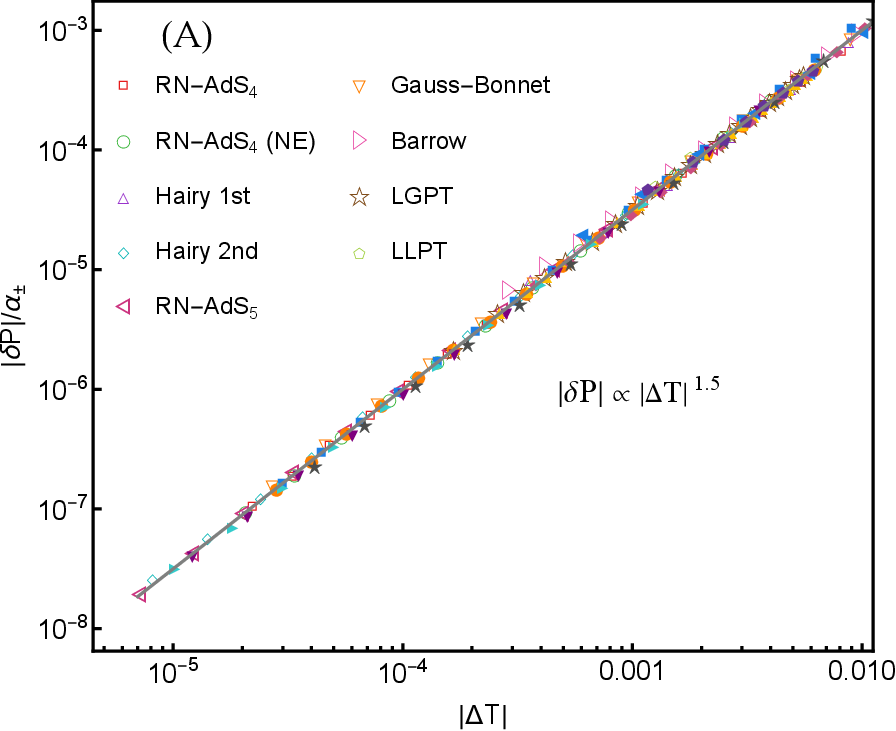}
    \includegraphics[width=0.48\columnwidth]{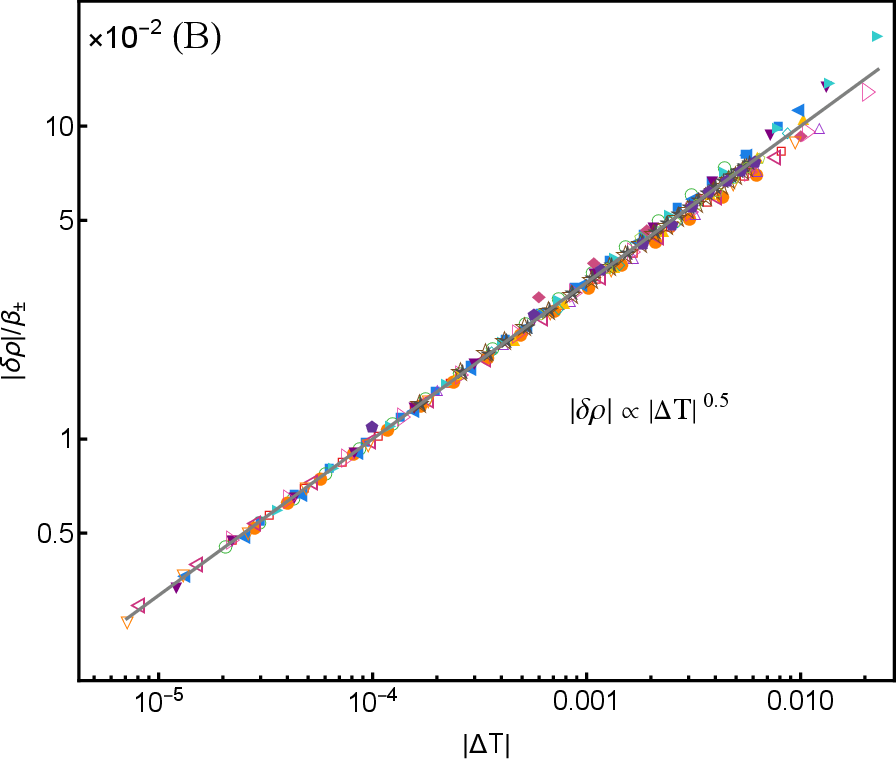}
    \caption{Scaling law near the critical point of (A) the external field, Eq.~(\ref{eq:H_scaling}), and (B) the order parameter, Eq.~(\ref{eq:eta_scaling}), along the supercritical crossover lines, $L^+$ and $L^-$ (open and filled symbols correspondingly),   for six black holes, the Van der Waals LGPT, and the LLPT in the two-state model.
    The values of $\alpha_\pm$ and $\beta_\pm$ are summarized in Table~\ref{table:coefficients}.
    }
    \label{fig:scalings} 
\end{figure}

Our numerical calculations confirm that near all critical points, the scaling of the $L^\pm$ lines follows Eqs.~(\ref{eq:H_scaling}) and~(\ref{eq:eta_scaling}). Although asymptotically exact only as $T\to T_c^+$, these scaling laws are found to hold effectively in a broad supercritical window for all systems studied, where higher-order corrections~\cite{li2026Supercritical} remain negligible. This is illustrated in Fig.~\ref{fig:scalings}, which includes fourteen supercritical  scaling lines derived from seven distinct black hole critical points (each critical point corresponds to two lines, $L^\pm$). As summarized in Table~\ref{tab:analogy}, for the extended phase-space (E-BH), $P$ is the external field and $\rho$ is the order parameter. For the non-extended phase space (NE-BH), $T$ serves as the external field, with $S$ being the corresponding order parameter ($S_B$ in Barrow black holes). Note that other definitions of supercritical crossover lines, including Widom~\cite{xu2005relation,Li:2025tdd} and Frenkel lines~\cite{brazhkin2012two, cockrell2021transition}, do not enjoy the universal scalings, Eqs.~(\ref{eq:H_scaling}) and~(\ref{eq:eta_scaling}).

\begin{table}[!htbp]
\centering
\caption{Analogies between charged black holes (in  the extended phase space (E-BH) and non-extended phase space (NE-BH)) and fluid systems (LGPT and LLPT). 
}
\begin{tabular}{ l  l l  l  l }
\hline
\hline
               & E-BH & NE-BH & LGPT & LLPT    \\
\hline
control field  & $T$ & $Q$ & $T$ & {$P$} \\
ordering field  & $P$ & $T$ & $P$ & {$T$} \\
order parameter & $\rho$ & $S$  & $\rho$ &  {$S$}\\
\hline 
\hline
\end{tabular}
\label{tab:analogy}
\end{table}

\subsection*{Analogous to liquid-gas phase transitions and liquid-liquid phase transitions in fluids}

The charged black hole EOSs in the extended thermodynamic phase space, given by Eq.~\eqref{eq:BHEPS_EOS}, have been compared to the Van der Waals EOS (see, \emph{e.g.}~\cite{Kubiznak:2012wp}). This analogy is illustrated by comparing the EOSs and phase diagrams of black holes (Fig.~\ref{fig:4DPV}A, B) with those of the LGPT in water (SM Fig.~S2A, B).
Extending this analogy to the supercritical regime, we find that both systems obey the same universal scaling laws along the $L^{\pm}$ lines, as given by Eqs.~(\ref{eq:H_scaling}) and~(\ref{eq:eta_scaling}) (see Fig.~\ref{fig:scalings} for the LGPT given by the Van der Waals EOS). 

The correspondence of standard charged black holes in the non-extended phase space to the liquid-gas system has been debated~\cite{Kubiznak:2012wp}. Instead, here we propose an analogy between the thermodynamics of these black holes and that of LLPTs. A LLPT is a first-order transition between low- and high-density liquid phases, terminated at a liquid-liquid critical point~\cite{poole1992phase}.
Comparing the EOSs and phase diagram of such black holes (Fig.~\ref{fig:4DTS}A, B) with those of a typical LLPT (SM Fig.~S2C, D) reveals two key similarities:
(i) The coexistence line in the phase diagram has a negative slope--a characteristic feature of LLPTs~\cite{luo2014behavior}---in contrast to the positive slope found in LGPTs.
(ii) A natural order parameter is the entropy \(S\), conjugate to the temperature \(T\). Notably, it has been suggested that LLPTs are entropy-driven rather than energy-driven~\cite{holten2012entropy}. Coincidently, Refs.~\cite{Chamblin:1999hg, Kubiznak:2012wp,HosseiniMansoori:2024bfi} have  proposed treating \(T\) as the ordering field and \(S\) as the order parameter for charged black holes in the non-extended phase space.
These correspondences are summarized in Table~\ref{tab:analogy}. We find that both black holes and LLPTs obey the same universal supercritical scaling laws given by Eqs.~(\ref{eq:H_scaling}) and~(\ref{eq:eta_scaling}) (see Fig.~\ref{fig:scalings}). EOSs and phase diagrams of the LGPT and LLPT are summarized in SM Sec.~S2.

\section{Conclusion}

The main contribution of this work is the systematic extension of the $L^\pm$ crossover lines to black hole thermodynamics. Unlike other supercritical diagnostics, they exhibit universal scaling behavior across diverse black hole models. Our findings not only underscore the universality of the supercritical thermodynamics for black holes but also reinforce the idea that supercritical phenomena near critical points are governed by universal principles, even in strongly curved spacetime geometries. We have numerically verified the universal scaling of the $L^\pm$ crossover lines across a wide range of black hole models with diverse physical parameters and phase structures (Fig.~\ref{fig:scalings}). Although the critical exponents in the models studied here take mean-field values, our universality claim is not about the mean-field exponents themselves, but rather about the fact that the $L^\pm$ scalings hold over a finite supercritical window for the entire crossover lines across diverse gravitational models with drastically different phase structures, including cases beyond the Van der Waals equation of state. This robustness is not guaranteed by standard near-critical mean-field theory. 

Therefore, this leaves many interesting directions, including, (i) generalization to complex black holes whose phase diagrams are with multiple critical points~\cite{Frassino:2014pha,Cai:2024eqa}, and non-AdS black holes~\cite{Altamirano:2014tva,Simovic:2018tdy,Haroon:2020vpr,Dayyani:2018mmm,Wang:2020hjw}, (ii) extension to non-mean-field theories by incorporating quantum effects, such as those predicted by holographic duality or loop quantum gravity~\cite{Hu:2024ldp,HosseiniMansoori:2024bfi}, and (iii) exploration of supercritical dynamic crossover phenomena to bridge the gap between equilibrium and non-equilibrium criticality~\cite{lv2024quantum}.
(iv) Moreover, based on the celebrated  AdS/CFT correspondence, generalization of the current approach can also be  used to infer the QCD phase diagram~\cite{sordi2024introducing}.
{
(v) Finally, by further integrating tools from statistical mechanics and quantum field theory, future research may unravel deeper connections between black hole microstates and macroscopic phase structures, providing valuable insights into the fundamental nature of quantum gravity.}
~\\

\section{Methods}
\subsection*{Determination of $L^\pm$ supercritical crossover lines in charged black holes}\label{sec:definition}

We take the 4D charged RN-AdS black holes  as  examples to explain how to determine $L^\pm$ lines. It can be straightforwardly generalized to other black holes.

\begin{figure}[h!]
\centering
\includegraphics[scale=0.62]{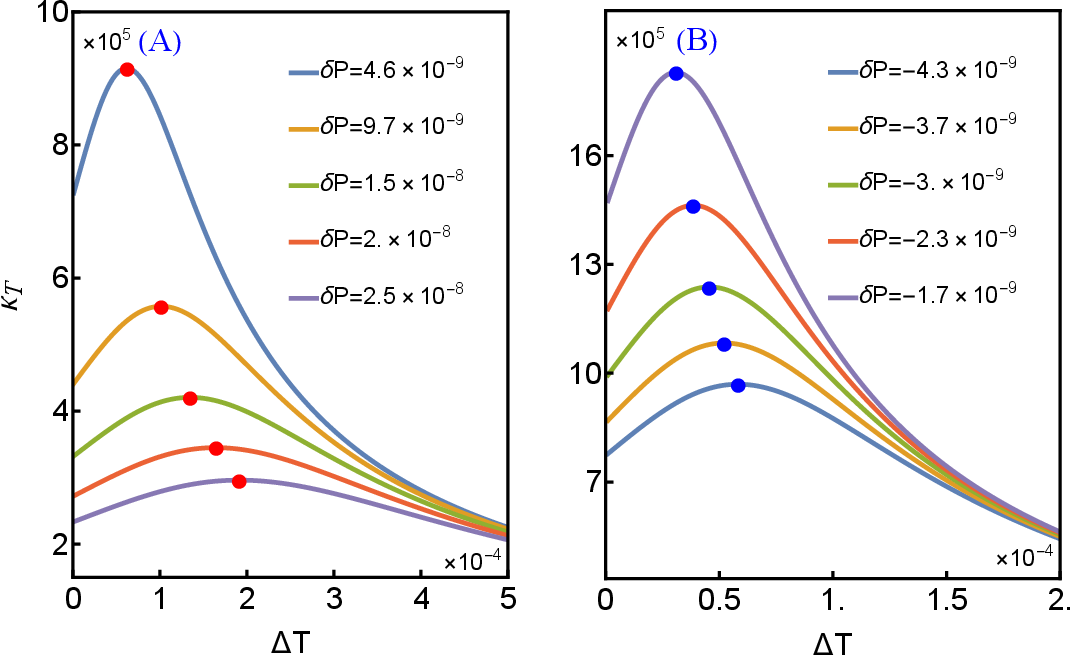}
\caption{Supercritical crossover lines for the RN-AdS black holes in the framework of extended phase space. The susceptibility $\kappa_T$ is shown as a function of $\Delta T$, for a few fixed (A) $\delta P>0$ and (B) $\delta P<0$. The peaks (red and blue dots) determine $L^{\pm}$. We have set $Q=1$.}
\label{fig:LEP} 
\end{figure}

We first focus on RN-AdS black hole in the extended phase space. To define the crossover lines proposed in~\cite{li2024thermodynamic}, one first chooses the critical isochore as an extension of the coexistence line to the supercritical region. The compressibility $\kappa_T$ is evaluated along each path parallel to the critical isochore. Since $\kappa_T$ is a function of distance $\delta P(v, T) = P(v,T) - P(v_c,T)$ and $T$, one can find a temperature $T_{max}(\delta P)$ that maximizes $\kappa_T$ along each path.
 All the $T_{max}(\delta P)$ points under different $\delta P$ together consist of the thermodynamic crossover lines $L^{\pm}$, on two sides of the critical isochore. Fig.~\ref{fig:LEP} depicts how the supercritical crossover lines $L^{\pm}$ are determined for the charged black hole in Fig.~\ref{fig:4DPV}. For a fixed $\delta P$, we find that $\kappa_T$ peaks at $T^+_{max}(\delta P)$ for a given $\delta P>0$ (Fig.~\ref{fig:LEP}A) and $T^-_{max}(\delta P)$ for a given $\delta P<0$ (Fig.~\ref{fig:LEP}B). The two resulting lines $L^{\pm}$ are shown explicitly in Fig.~\ref{fig:4DPV}B in the $(\Delta P,\Delta T$)-plane, where $\Delta P=P/P_c-1$ and $\Delta T=T/T_c-1$.\\ 

\begin{figure}[h!]
    \centering
    \includegraphics[scale=0.62]{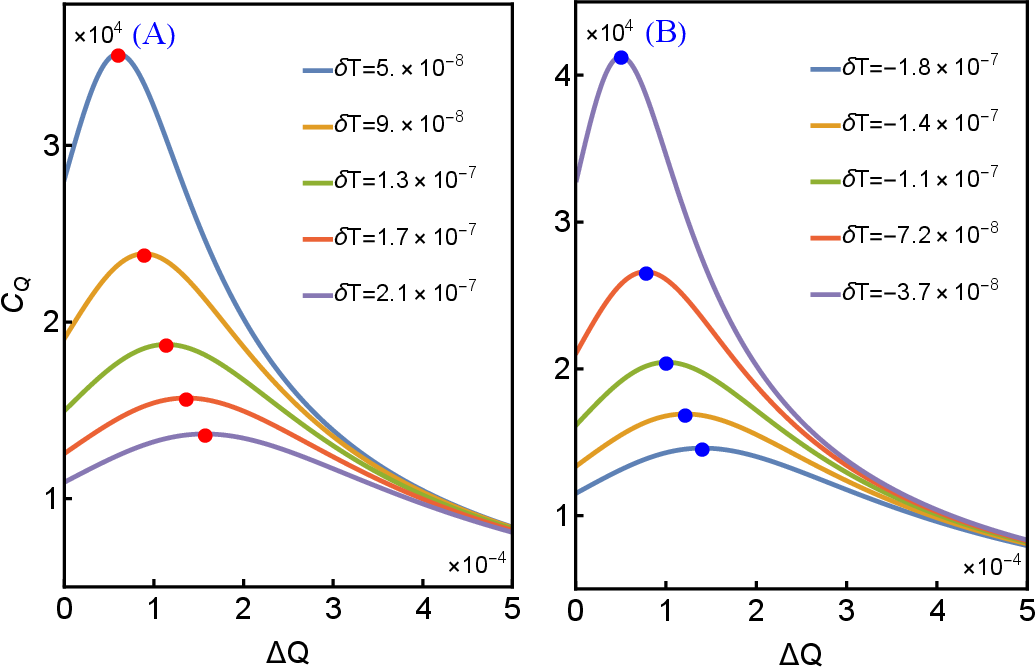}
    \caption{Supercritical crossover lines for the RN-AdS black
holes in the non-extended phase space. The specific heat $C_Q$ as a function of $\Delta Q$ for a few fixed $\delta T>0$ (A) and $\delta T<0$ (B), respectively. The peaks (red and blue dots) correspond to supercritical crossover lines $L^{\pm}$. We set $l=\sqrt{3}$.}
    \label{fig:TQcrossline} 
\end{figure}

{For the case in the non-extended framework, we consider $C_Q$ as a function of $\Delta Q=Q/Q_c-1$ for a few fixed $\delta T(S, Q)=T(S, Q)-T(S_c, Q)$. As shown in Fig.~\ref{fig:TQcrossline}A and B, for a fixed $\delta T$, the specific heat $C_Q$ peaks at $Q_{max}^+$ for $\delta T>0$ and peaks at $Q_{max}^-$ for $\delta T<0$. We then obtain the supercritical crossover lines $L^{\pm}$ shown explicitly in Fig.~\ref{fig:4DTS}B.}

\subsection*{Coefficients of $L^\pm$ lines}\label{sec:coefficients}

In Table~\ref{table:coefficients}, we list the  coefficients $\alpha_\pm$ and $\beta_\pm$ obtained from the power-law fitting  to the scaling forms, $\delta P^{\pm} =  \alpha_\pm (T-T_{\rm c})^{\beta + \gamma}$ and $\delta \rho^{\pm} =  \beta_\pm (T-T_{\rm c})^{\beta}$, for all systems in Fig.~\ref{fig:scalings} of main text.

\begin{table}[!htbp]
\centering
\caption{Values of the coefficients $\alpha_\pm$ and $\beta_\pm$. }
\begin{tabular}{lrrrr}
\hline
\hline
\qquad \qquad & $ \alpha_+$ & $   \alpha_-  $ & $\beta_+$ & $\beta_- $\\
\hline
RN-$\text{AdS}_4$ & 0.01  & 0.01 & 0.813 & 0.821 \\
RN-$\text{AdS}_4$ (NE)  & 0.106 & 0.105 & 1.43 & 1.39 \\
LGPT  &  9.41 & 9.04 & 1.14 & 1.17\\
LLPT  & 0.004 & 0.005 & 0.177 & 0.183 \\
Hairy 1st  & 0.075 & 0.094 & 1.08 & 1.12 \\
Hairy 2nd & 1.4 & 1.01 & 0.142 & 0.108 \\
Barrow  & 0.006 & 0.008 & 0.73 & 0.958 \\
RN-$\text{AdS}_5$  &  0.038 & 0.046 & 0.581 & 0.701\\
Gauss-Bonnet  & 0.052 & 0.058 & 1.14 & 1.36 \\
\hline
\hline
\end{tabular}
\label{table:coefficients}
\end{table}

\section*{Acknowledgments}
\textbf{Funding: }This work was partly supported by the National Natural Science Foundation of China Grants No.\,12525503, No.\,12588101, and 12447101. Y.J. acknowledges funding from   the National Key R\&D Program of China (Grant No. 2025YFF0512000), the Space Application System of China Manned Space Program (Grant No. CMSS-2025-5-P-002), and Wenzhou Institute (Grant WIUCASICTP2022). X. L.  acknowledges the Postdoctoral Fellowship Program of CPSF (Grant Number GZC20252776). The authors acknowledge the use of the High Performance Cluster at Institute of Theoretical Physics, Chinese Academy of Sciences,  and the computer clusters at the Hefei advanced computing center.\\
\textbf{Author contributions: }S.W.: Theoretical derivation and numerical simulation. X.L.: Methodology and data analysis. Y.J.: Design and interpretation, data analysis. L.L.: Conceptualization, methodology, design and interpretation. \\
\textbf{Competing interests: }The authors declare that they have no competing interests.

\section*{Data Availability}
The data that support the findings of this study are available from the corresponding author upon reasonable request. 

\section*{Supplementary Materials}

\subsection*{S1. Black hole models}\label{sec:bhmodels}

We provide detailed description of the black hole models discussed in the main text. Except for 4D RN-AdS black holes, we consider the extended phase space in the canonical ensemble.
The geometric units, $G_N=\hbar=c=k_B=1$, are used.

\subsubsection*{A. The RN black hole in $\text{AdS}_4$}

We begin with the spherical RN-AdS black hole in 4-dimensional spacetime from the Einstein-Maxwell action,
\begin{equation}
    S_{EM}=\frac{1}{16\pi}\int d^4x\sqrt{-g}\left( R-F_{\mu\nu}F^{\mu\nu}-2\Lambda \right)\,,
\end{equation}
where $g$ is the determinant of the metric $g_{\mu\nu}$, $R$ the corresponding Ricci scalar, $F_{\mu\nu}=\partial_\mu A_\nu-\partial_\nu A_\mu$ the field strength of the U(1) gauge field $A_\mu$, and $\Lambda=-3/l^2$ with $l$ the AdS radius. The black hole solution reads
\begin{equation}
    ds^2=-f(r)dt^2+\frac{dr^2}{f(r)}+r^2(d\theta^2+\sin{\theta}^2d\phi^2)\,, \label{AdS4metric}
\end{equation}
where
\begin{equation}
    f(r)=1-\frac{2M}{r}+\frac{Q^2}{r^2}+\frac{r^2}{l^2}\,,
\end{equation}
and the gauge potential $A_t=-Q/r$. The parameter $M$ represents the ADM mass of the black hole and $Q$ the total charge. 

The black hole event horizon $r_h$ is determined as a larger root of $f(r_h)=0$. The Hawking temperature and entropy are given by  $T=\frac{1}{4\pi r_h}\left ( 1+\frac{3r_{h}^{2} }{l^2}-\frac{Q^2}{r_h^2}   \right )$ and  $S=\pi r_h^2$, respectively.
%
%
The chemical potential is given by $\mu=Q/r_h$,  which measures the potential difference between the horizon and infinity. For example, one can obtain the specific heat $C_Q \equiv T\frac{\partial S}{\partial T}|_Q=\frac{2 \pi  r_h^2 \left(-l^2 Q^2+l^2 r_h^2+3 r_h^4\right)}{3 l^2 Q^2-l^2 r_h^2+3 r_h^4}$. Moreover, in the extended phase
space, the thermodynamic volume reads $V=4\pi r_h^3/3$ and the specific volume is given by $v=2r_h$.\\

\subsubsection*{B. The RN black hole in $\text{AdS}_5$}
The natural extension of the above-disuccsed black hole is to consider higher dimensions. To illustrate this, we investigate a five-dimensional spherical RN-AdS black hole as a representative case. The geometry is given by~\cite{Niu:2011tb}
\begin{equation}
\begin{split}
ds^2 = -f(r)dt^2 + \frac{dr^2}{f(r)} + r^2 d\Omega_3^2\,,\\
f(r) = 1 - \frac{8M}{3\pi r^2} + \frac{Q^2}{r^4} + \frac{r^2}{l^2}\,,
\end{split}
\end{equation}
with $d\Omega_3^2$ the metric of 3-sphere. The thermodynamic pressure in the extended phase space is defined as
\begin{equation}
P = -\frac{\Lambda}{8\pi} = \frac{3}{4\pi l^2}\,.
\end{equation}
In the canonical ensemble, one has the first law of black hole thermodynamics:
\begin{equation}
dM = TdS + \mu dQ + VdP\,,
\end{equation}
where $V=\pi^2 r_h^4/2$ is the thermodynamic volume conjugate to $P$ with $r_h$ the location of the event horizon.

Using the `Euclidean trick', one can obtain the black hole temperature
\begin{equation}
T = \frac{f'(r_h)}{4\pi} = \frac{1}{2\pi r_h} \left( 1 + \frac{2r_h^2}{l^2} - \frac{Q^2}{r_h^4} \right),
\end{equation}
and the pressure
\begin{equation}
P = \frac{3}{8\pi r_h^2} \left( 2\pi T r_h - 1 + \frac{Q^2}{r_h^4} \right)\,.
\end{equation}
By identifying the specific volume as $v = 2r_h$, one obtains the equation of state
\begin{equation}
P = \frac{3T}{2v} - \frac{3}{2\pi v^2} + \frac{24Q^2}{\pi v^6}.
\end{equation}
It closely resembles the 4-dimensional RN black hole, except for the power of the last term in $v$.

The critical point is determined by
\begin{align}
\left(\frac{\partial P}{\partial v}\right)_{T,Q} = 0,\quad 
\left(\frac{\partial^2 P}{\partial v^2}\right)_{T,Q} = 0\,,
\label{eq:critical_conditions}
\end{align}
from which one obtains
\begin{equation}
T_c = \frac{2 \sqrt{2}}{5 \sqrt[4]{15} \pi  \sqrt[4]{Q^2}},\;  v_c = (60Q^2)^{1/4}, \; P_c = \frac{1}{6 \sqrt{15} \pi  Q}\,.
\end{equation}
To compute the critical exponents for the critical point, we introduce
\begin{align}
t= \frac{T - T_c}{T_c},\quad \omega = \frac{v - v_c}{v_c}\,.\label{definetandw}
\end{align}
This notation is not to be confused with the temporal coordinate $t$.
Then one obtains the equation of state near the critical point:
\begin{equation}
   \hat{P}\equiv \frac{P}{P_c}=1 + \frac{12}{5}t - \frac{12}{5}t\omega - 2\omega^3 + O(\omega^4, t\omega^2)\,.
  \label{eq:expansion}
\end{equation}
We are interested in the following two critical exponents. The critical exponent $\beta$ describes the behavior of the order parameter:
\begin{equation}
\eta = v_l - v_s \propto |t|^\beta\,.
\end{equation}
with $v_s$ and $v_l$ the specific volumes for small and large black holes.
The second critical exponent $\gamma$ describes the divergence of isothermal compressibility:
\begin{equation}
\kappa_T = -\frac{1}{v}\left(\frac{\partial v}{\partial P}\right)_T \propto |t|^{-\gamma}\,.
\end{equation}

For an isotherm at temperature $t<0$, there is a first-order phase transition. At the transition point, both phases possess the same Gibbs free energy, in accordance with Maxwell's equal-area rule:
\begin{equation}\label{Maxwelllaw}
 \oint v dP=0\,,  
\end{equation}
which replaces the ‘oscillating part’ of the isotherm by an isobar. Applying~\eqref{Maxwelllaw} for fixed $t < 0$, one has
\begin{equation}
\int_{\omega_l}^{\omega_s} \omega \left(\frac{\partial \hat{P}}{\partial \omega}\right)_t d\omega=\int_{\omega_l}^{\omega_s} \omega\left(-\frac{12}{5}t - 6\omega^2\right) d\omega  = 0\,,
\end{equation}
from which 
\begin{equation}
\left[-\frac{6}{5}t\omega^2 - \frac{3}{2}\omega^4\right]_{\omega_l}^{\omega_s} = 0\,.
\label{eq:area_law}
\end{equation}
Moreover, during the phase transition the pressure remains constant:
\begin{equation}
\hat{P}(\omega_s) = \hat{P}(\omega_l)\,.
\end{equation}
Then from~\eqref{eq:expansion} one has
\begin{equation}
-\frac{12}{5}t(\omega_s - \omega_l) - 2(\omega_s^3 - \omega_l^3) = 0\,.
\label{RN:pressure_equality}
\end{equation}
The two equations~\eqref{eq:area_law} and~\eqref{RN:pressure_equality} have a unique non-trivial solution $\omega_s = -\omega_l =\sqrt{6/5} \sqrt{-t}$.
Therefore, one has the order parameter given by
\begin{equation}
\eta = v_c(\omega_l - \omega_s) =  2v_c\sqrt{-\frac{6}{5}t} \propto \sqrt{-t}\Rightarrow \beta = \frac{1}{2}\,.
\end{equation}

Moreover, from (\ref{eq:expansion}), one has
\begin{equation}
\left(\frac{\partial \hat{P}}{\partial \omega}\right)_t = -\frac{12}{5}t - 6\omega^2\,.
\end{equation}
Then, along the critical isochore ($\omega = 0$), it becomes
\begin{equation}
\left(\frac{\partial \hat{P}}{\partial \omega}\right)_t = -\frac{12}{5}t \quad \Rightarrow \quad \left(\frac{\partial \omega}{\partial \hat{P}}\right)_t = -\frac{5}{12t}\,.
\end{equation}
Hence, the isothermal compressibility scales as
\begin{equation}
\kappa_T \propto \left(\frac{\partial v}{\partial P}\right)_T = \frac{v_c}{P_c}\left(\frac{\partial \omega}{\partial \hat{P}}\right)_t \propto \frac{1}{t}\Rightarrow \gamma=1\,.
\end{equation}

\subsubsection*{C. Hairy electrically charged black hole}
This model is an exact hairy black hole constructed by RN-AdS model with a non-trivial self-interacting scalar potential. The action is given by~\cite{Astefanesei:2019ehu}:
\begin{equation}
\begin{split}
   I\left[g_{\mu\nu},A_{\mu},\phi \right]&=\frac{1}{16\pi}\int_{\mathcal{M}}d^{4}x\sqrt{-g}\left[R-e^{\phi}F_{\mu\nu}F^{\mu\nu}\right.\\&\left.-\frac{1}{2}\left(\partial\phi \right)^{2} -V(\phi) \right],
\end{split}
\end{equation}
where the dilaton potential takes the form:
\begin{equation}
V(\phi)=\left(\frac{\Lambda}{3}+\phi\right)(4+2\cosh\phi)-6\sinh\phi\,.
\end{equation}
The hairy black hole solution is given by
\begin{align}
ds^2 &= \Omega(x) \left[ -f(x)dt^2 + \frac{\eta^2 dx^2}{x^2 f(x)} + d\theta^2 + \sin^2 \theta d\phi^2 \right]\,, \label{eq:metric} \\
A_t &= \left( \frac{q}{x} - C \right), \quad \phi = \ln(x) \,,\label{eq:potential}
\end{align}
where \( C \) is a constant and the explicit expressions for the conformal factor \(\Omega(x)\) and metric function \( f(x) \) are
\begin{equation}
\begin{split}
    &\Omega(x) = \frac{x}{\zeta^{2} (x-1)^{2}}\,,\\
&f(x) = \frac{x^{2} - 1}{2x} - \ln x 
+ \frac{\zeta^{2} (x-1)^{2}}{x} \left( 1 - \frac{2q^{2}}{x} \right) 
- \frac{\Lambda}{3}\,.
\end{split}
\end{equation}
At the event horizon $x_h$, one has $f(x_h)=0$. The constant $C$ in the gauge potential is fixed by $A_t(x_h)=0$. We focus on the branch that corresponds to thermodynamically stable black holes in flat space in the limit
$\Lambda=0$. 

In the extended framework with $P=-\Lambda/(8\pi)$, As shown in~\cite{Astefanesei:2019ehu}, the system
in the \textit{canonical ensemble} (fixed charge $Q$) shows \textit{two distinct critical points}, unlike the single critical point in the RN-AdS case. We now derive the critical exponents $\beta$ and $\gamma$ for the hairy black holes in the canonical ensemble.

The temperature is given by
\begin{equation}
T = \frac{(x_h - 1)^2}{8\pi \zeta x_h} \left[ -1 - 2\zeta^2 \left( \frac{x_h + 1}{x_h - 1} \right) + 4\zeta^4 Q^2 \left( \frac{x_h + 2}{x_h} \right) \right]\,,
\label{eq:temperature}
\end{equation}
and the specific volume reads
\begin{equation}
v = \frac{1}{\zeta} \left( \frac{x_h + 1}{x_h - 1} \right)\,.
\label{eq:specific_volume}
\end{equation}
with the parameter $\zeta$ determined from the horizon equation
\begin{equation}
\begin{split}
\zeta & = \frac{1}{2Q} \sqrt{\frac{x_h}{x_h - 1}} \left[ 1 + \right.\\&\left. \sqrt{1 +  \frac{4(16\pi x_h P - 6x_h \ln x_h + 3x_h^2 - 3)Q^2}{3x_h(x_h - 1)}} \right]^{1/2}\,.
\label{eq:eta_relation}
\end{split}
\end{equation}
Unlike the AdS-RN black hole, it is not possible to get an analytic expression $P=P(v, T, Q)$ for the hairy case. Nevertheless, one has the same criticality conditions~\eqref{eq:critical_conditions}. There are two critical points for any value of $Q$, including a RN-AdS–like critical point at large volume and low temperature, and another critical point at small volume and high temperature. As
$Q$ is varied, the critical compressibility factor $Z_c=P_c v_c/T_c$ changes continuously, effectively tuning the nature of the fluid in dual theory.

For a fixed $Q$, we introduce the expansion variables $t$ and $\omega$ near each critical point, following the definition given in~\eqref{definetandw}.
Near the critical point, the equation of state can then be expanded as:
\begin{equation}
P = P_c + A t + B t \omega + C \omega^3 + \mathcal{O}(t\omega^2, \omega^4)\,,
\label{eq:pressure_expansion}
\end{equation}
following from the criticality  conditions \eqref{eq:critical_conditions}. Here the parameters $A, B, C$ depend solely on $Q$.

Differentiating equation \eqref{eq:pressure_expansion} at fixed $t < 0$, one has
\begin{equation}
dP = B t d\omega + 3C \omega^2 d\omega\,.
\label{eq:pressure_diff}
\end{equation}
Let $\omega_s$ and $\omega_l$ denote the specific volume deviations for small and large black holes, respectively. The pressure equality condition gives
\begin{equation}
P_c + A t + B t \omega_s + C \omega_s^3 = P_c + A t + B t \omega_l + C \omega_l^3\,,
\end{equation}
which simplifies to
\begin{equation}
B t (\omega_s - \omega_l) + C (\omega_s^3 - \omega_l^3) = 0\,.
\label{eq:pressure_equality}
\end{equation}

Note that the Maxwell's equal area law~\eqref{Maxwelllaw} requires
\begin{equation}
\int_{\omega_s}^{\omega_l} \omega (B t + 3C \omega^2) d\omega = 0\,.
\label{eq:maxwell_law}
\end{equation}
By solving equations \eqref{eq:pressure_equality} and \eqref{eq:maxwell_law}, we have $\omega_l = -\omega_s = \omega_0$. Then one obtains from~\eqref{eq:pressure_equality} that
\begin{equation}
B t + C \omega_0^2 = 0\,, \quad \Rightarrow \quad \omega_0^2 = -\frac{B}{C} t\,.
\label{eq:omega_relation}
\end{equation}
The order parameter (volume difference between phases) is given as
\begin{equation}
\eta = v_c (\omega_l - \omega_s) = 2v_c \sqrt{-\frac{B}{C}} \sqrt{-t} \propto \sqrt{-t}\Rightarrow \beta = \frac{1}{2}\,.
\end{equation}

Furthermore, from~\eqref{eq:pressure_expansion}, the pressure derivative is
\begin{equation}
\left( \frac{\partial P}{\partial v} \right)_{T} = B t + 3C \omega^2\,.
\label{eq:pressure_deriv}
\end{equation}
Near the critical point ($\omega \to 0$), one has
\begin{equation}
\left( \frac{\partial P}{\partial v} \right)_{T} \approx B t\,.
\end{equation}
Thus, the isothermal compressibility behaves as 
\begin{equation}
\kappa_T = -\frac{1}{v} \left( \frac{\partial v}{\partial P} \right)_{T} \approx -\frac{1}{v_c} \cdot \frac{1}{B t} \propto \frac{1}{t}\Rightarrow \gamma = 1\,.
\end{equation}

\subsubsection*{D. Gauss-Bonnet Black Hole}

We consider the Gauss–Bonnet black hole in AdS$_5$ space, incorporating higher-curvature corrections. The action in five-dimensional spacetime is given by~\cite{Cai:2013qga}:

\begin{equation}
S = \frac{1}{16\pi} \int d^5 x \sqrt{-g} \left[ R - 2\Lambda + \alpha_{\text{GB}} R_{\text{GB}} - F_{\mu\nu} F^{\mu\nu} \right],
\end{equation}
where $R_{\text{GB}} = R_{\mu\nu\gamma\delta} R^{\mu\nu\gamma\delta} - 4 R_{\mu\nu} R^{\mu\nu} + R^2$. The charged black hole with spherical symmetric horizon reads
\begin{equation}
 ds^2=-f(r)dt^2+f^{-1}(r)dr^2+r^2d\Omega_3^2\,,   
\end{equation}
where
\begin{equation}
\begin{split}
f(r)=&1+\frac{r^2}{4\alpha_{\text{GB}}}\left(1- \right.\\&\left. \sqrt{1+\frac{64\alpha_{\text{GB}}M}{3\pi r^4}-\frac{2\alpha_{\text{GB}}Q^2}{3r^6}-\frac{32\pi \alpha_{\text{GB}}P}{3}}\right)\,,
\end{split}
\end{equation}
and from which the thermodynamic quantities are
\begin{align}
M &= \frac{\pi}{8} \left(4\pi P r_h^4 + 3r_h^2 + 6\alpha_{\text{GB}} + \frac{Q^2}{4\pi r_h^2}\right)\,, \\
T &= \frac{8\pi P r_h^3 + 3r_h - \frac{Q^2}{8\pi r_h^3}}{2\pi (r_h^3 + 8\alpha_{\text{GB}} r_h)}\,, \\
S &= \frac{\pi^2}{2} \left( r_h^{3} + 12\alpha_{\text{GB}} r_h \right), \\
P &= -\frac{\Lambda}{8\pi}=\frac{3}{4r_h} \left( 1 + \frac{4\alpha_{\text{GB}}}{r_h^2} \right) T - \frac{3}{8\pi r_h^2} + \frac{Q^2}{8\pi r_h^6}\,.
\end{align}

This model exhibits Van der Waals-type $P$--$V$ criticality and small/large black hole phase transitions. Following a computational strategy similar to that used in previous sections, we obtain the critical exponents $\beta = \frac{1}{2}$ and $\gamma = 1$.

\subsubsection*{E. Charged black hole in the Barrow model}
The Barrow model incorporates quantum gravitational effects through a quasi-fractal structure of spacetime, often referred to as spacetime foam. In this approach, the black hole entropy is modified according to the relation proposed by Barrow~\cite{Barrow:2020tzx},
\begin{equation}
S_B = \left( \frac{A}{4G} \right)^{1+\delta/2},
\end{equation}
where $\delta$ is the Barrow parameter ($0 \leq \delta \leq 1$) quantifying the deviation from a smooth horizon geometry. Here $\delta = 0$ corresponds to the standard Bekenstein–Hawking entropy, and $\delta = 1$ corresponds to the maximal fractal influence.

Within this framework, the metric for a charged AdS black hole is given by \eqref{AdS4metric}. Nevertheless, the thermodynamic quantities are modified by the Barrow parameter $\delta$:
\begin{align}
S &= \left(\pi r_h^2\right)^{1+\frac{\delta}{2}}\,, \\
T &= \frac{8\pi P r_h^4 + r_h^2 - Q^2}{2\pi^{1+\frac{\delta}{2}}(\delta+2) r_h^{3+\delta}}\,, \\
P &= \frac{3}{8\pi \ell^2}= \frac{(1 + \delta)^2 \left( \pi r_h^2 \right)^{\delta/2} T}{4r_h} + \frac{Q^2}{8\pi r_h^4} - \frac{1}{8\pi r_h^2}\,, \\
V &= \frac{4}{3}\pi r_h^3\,.
\end{align}
The modified Smarr relation becomes
\begin{equation}
M = (2+\delta)TS + \mu Q - 2PV\,.
\end{equation}
Adopting a similar computational strategy, we calculate the critical exponents, yielding $\beta = \frac{1}{2}$ and $\gamma = 1$.

\begin{figure*}[hpt]
    \centering
    \includegraphics[width=0.315\columnwidth]{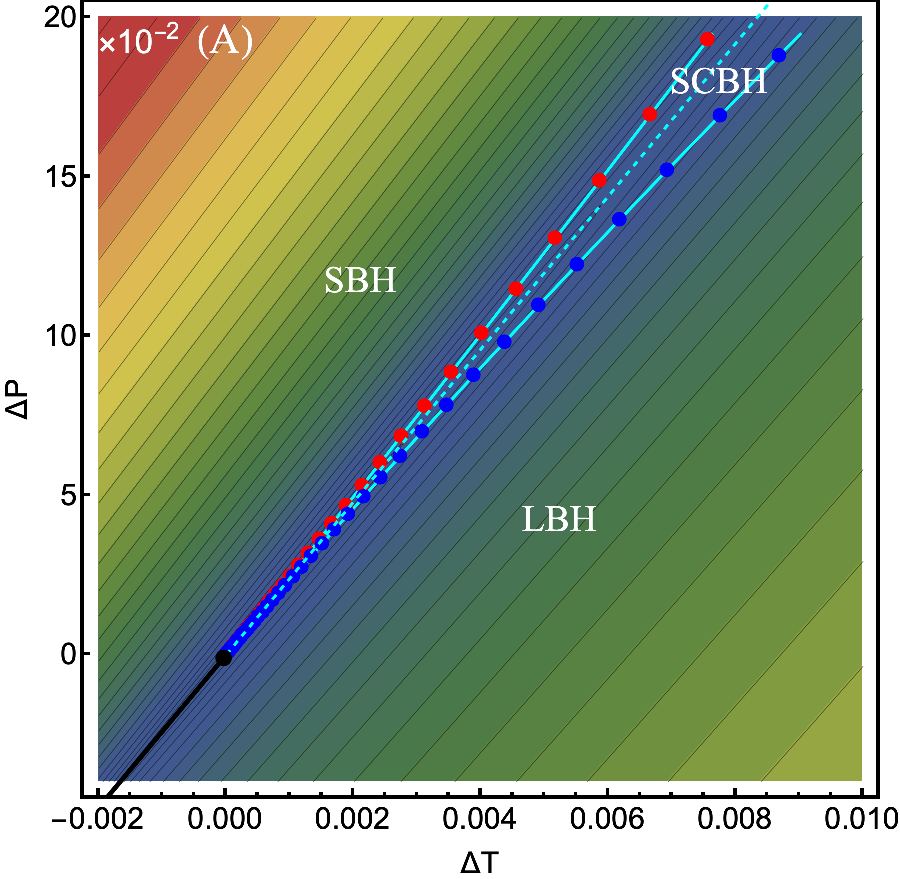}
    \includegraphics[width=0.31\columnwidth]{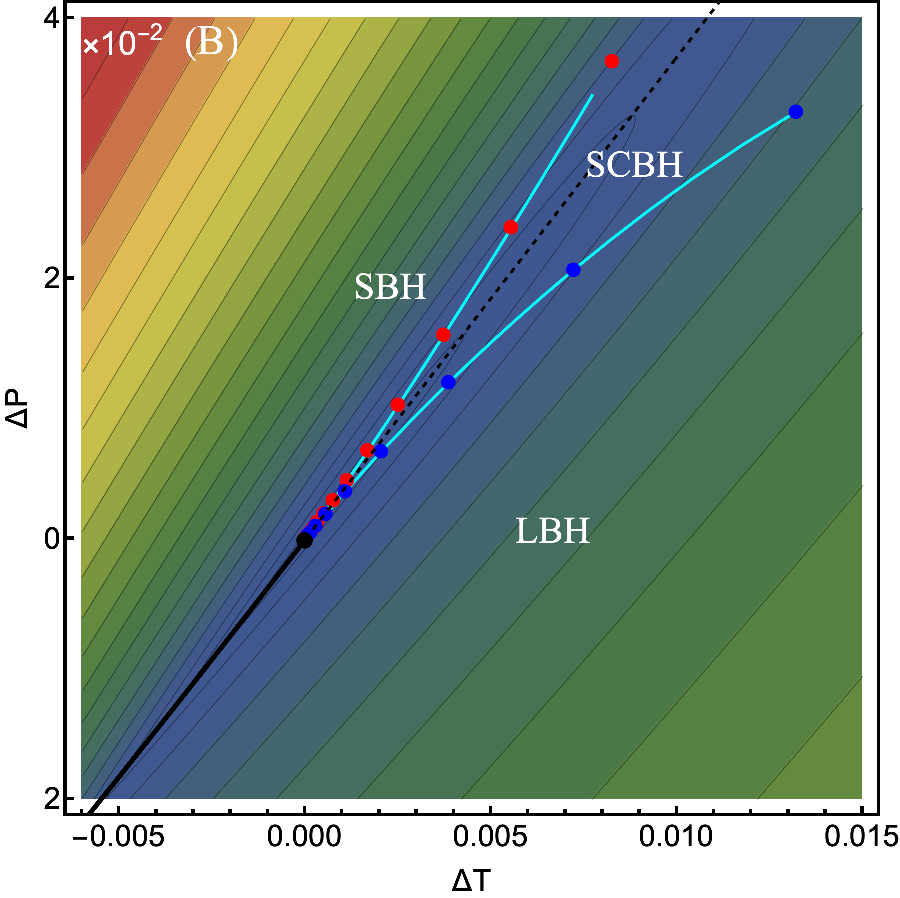}
    \includegraphics[width=0.35\columnwidth]{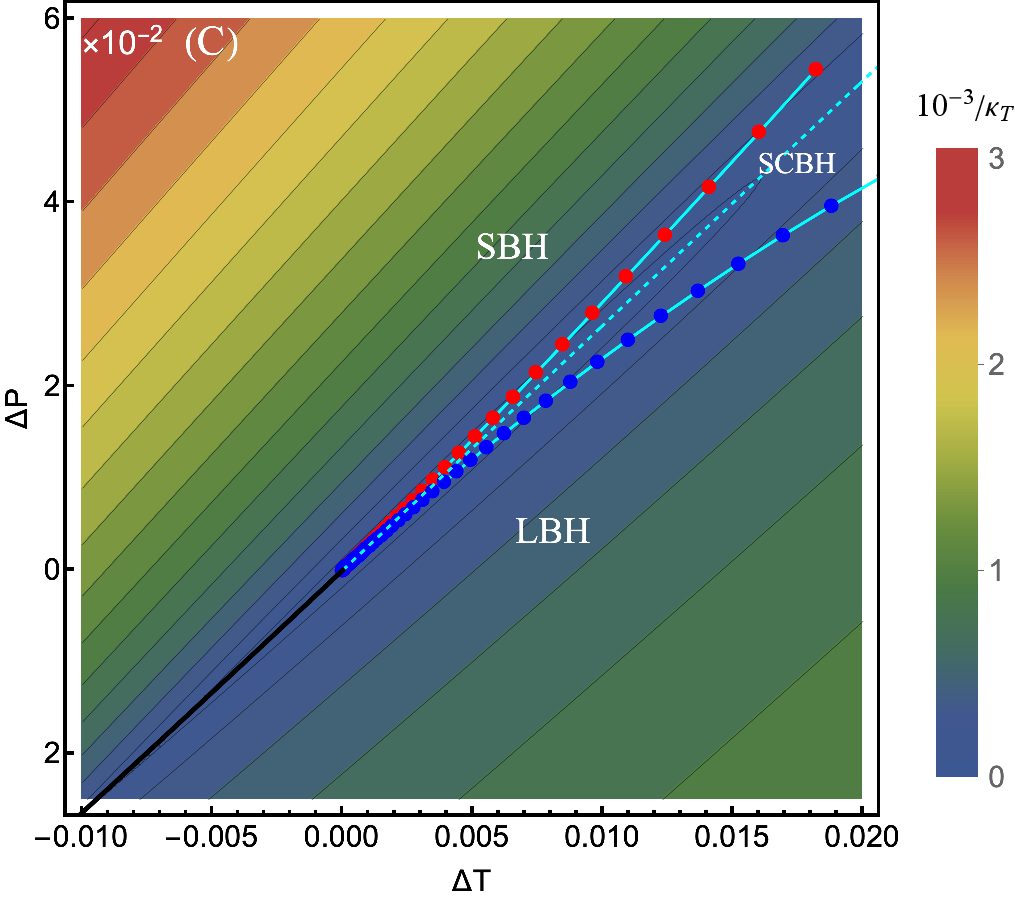}
    \caption*{Fig. S1: Phase diagrams  corresponding to 
    (A) $\text{RN-AdS}_5$ black hole, 
    (B) Gauss-Bonnet black hole in $\text{AdS}_5$ space, and
    (C) Charged AdS Black Hole of Barrow model. In all three systems, the charge is set to $Q=1$. For the Gauss-Bonnet black hole we set $\alpha_{\text{GB}}=1/2$, and for the Barrow-model black hole we set $\delta=0.8$.
 }
    \label{fig:PTcontour} 
\end{figure*}

\subsubsection*{F. Supercritical crossover lines $L^\pm$}
For each model, we have derived the critical exponents \(\beta\) and \(\gamma\) at their respective critical points, confirming their adherence to mean-field values. Subsequently, we numerically determine the supercritical crossover lines \(L^{\pm}\) following the methodology outlined in the End Matter. The scaling behaviors of these lines, governed by Eqs.~(2) and (3) in the main text, are fully verified. The results, as exemplified in Fig.~4 of main text for selected critical points (see Fig.~S1 for phase diagrams), demonstrate that the proposed scalings hold universally across these diverse and complex black hole backgrounds, reinforcing the mean-field nature of their criticality and the generality of our findings.
We point out that the hairy electrically charged black hole with two critical points are clearly beyond the mean-field Van der Waals equation of state.

\subsection*{S2. EOSs and phase diagrams of the LGPT and LLPT}\label{sec:lgptandllpt}

To study the supercritical thermodynamics near a LGPT, we collect the EOS of water, $\hat{P}(\hat{\rho}; \hat{T})$, in the supercritical regime from the NIST database~\cite{NIST}.
The thermodynamic variables are rescaled by their critical values, including temperature $\hat{T} = T/T_{\rm c}$, pressure $\hat{P} = P/P_{\rm c}$ and density $\hat{\rho} = \rho/\rho_{\rm c}$, {where $T_{\rm c} = 647.10~\rm{K}$, $P_{\rm c}=22.06~\rm{MPa}$
 and $\rho_{\rm c} = 17.87~\rm{mol/l}$}.
The pressure $P$, temperature $T$, and density $\rho$ are regarded respectively as the external field, control field, and order parameter (see Table~1 in main text).   The corresponding response function is the  susceptibility (compressibility) $\beta_T  =  \left. \frac{P_{\rm c}}{\rho_{\rm c}}  \frac{\partial \rho}{\partial  P} \right|_T$. The $L^\pm$ lines are determined by finding the maximum $\beta_T$ along paths parallel to the critical isochore. Fig.~S2(A,B) shows typical EOSs and the phase diagram near the LGPT, to be compared with those for 4D RN-Ads black holes in the extended phase space (Fig.~1A,B in main text).

To study the supercritical thermodynamics near a LLPT,
we consider a generic mean-field two-state liquid-liquid model~\cite{bertrand2011peculiar,  holten2012entropy} (see Sec.~S3A for details). For a given system, the model can be used to fit the experimental data. The model parameters are taken from~\cite{ holten2012entropy} such that the EOS given by the model agrees quantitatively  with the available experimental data of supercooled water near the conjectured  low-density liquid (LDL) to high-density liquid (HDL) phase transitions and the associated critical point.
As summarized in Table~1 of main text, the order parameter of the LLPT is  $S$, since the transition is entropy-driven. The typical  EOSs and the phase diagram of the two-state model are plotted in Fig.~S2C,D, to be compared with those for 4D RN-Ads black holes in the non-extended phase space (Fig.~2 in main text). Since the existence of LLPTs in water is still under debate~\cite{gallo2016water}, we further analyze a linear scaling theory with a negatively sloped ($dP/dT < 0$) coexistence line, which is a well-known theoretical model for general LLPTs  (see Sec.~S3B)~\cite{luo2014behavior}. The robustness of the universal supercritical scalings is confirmed. 

\begin{figure*}[t!]
    \centering
    \includegraphics[width=0.4\columnwidth]{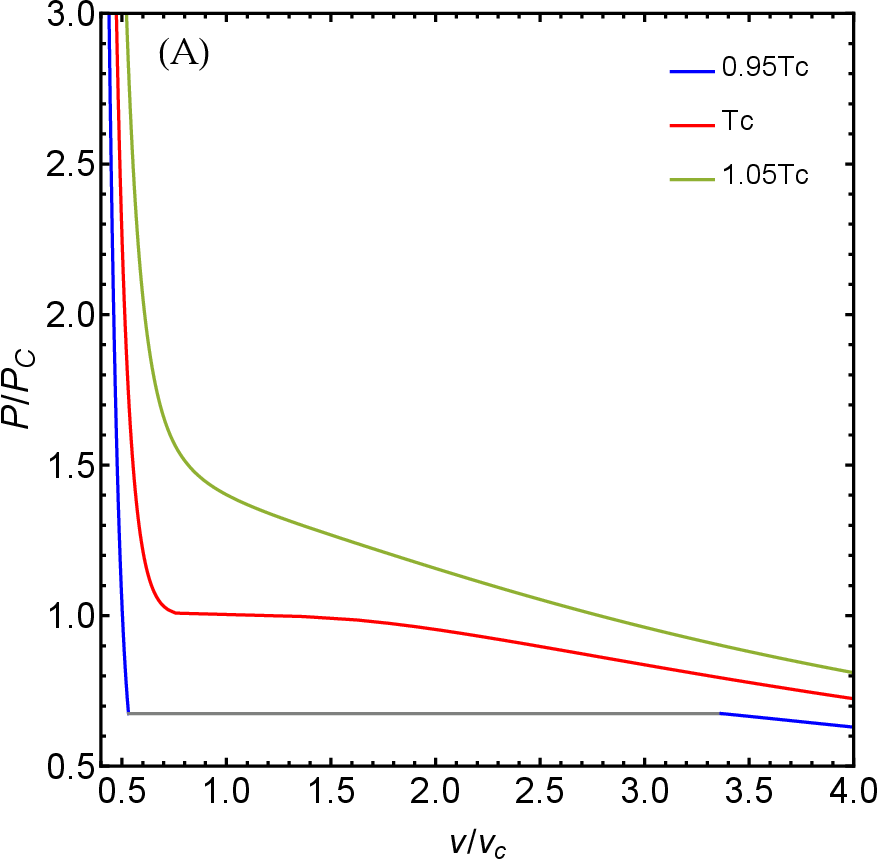}
    \includegraphics[width=0.43\columnwidth]{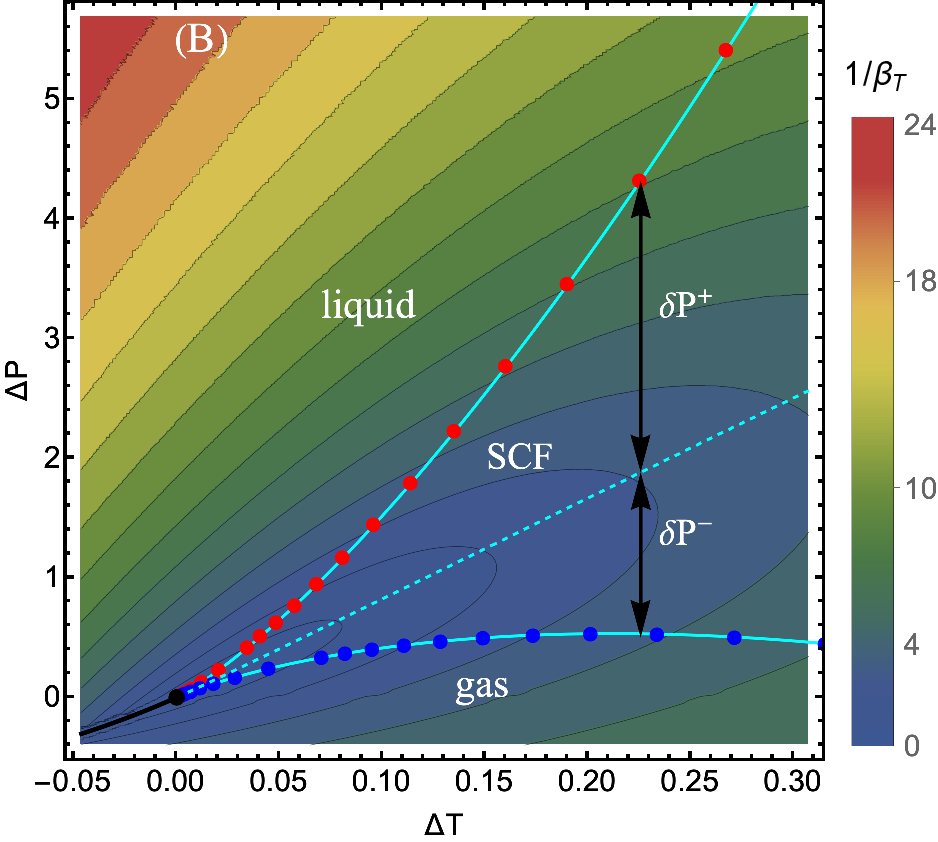}
    \hspace*{-0.3cm}
    \includegraphics[width=0.4\columnwidth]{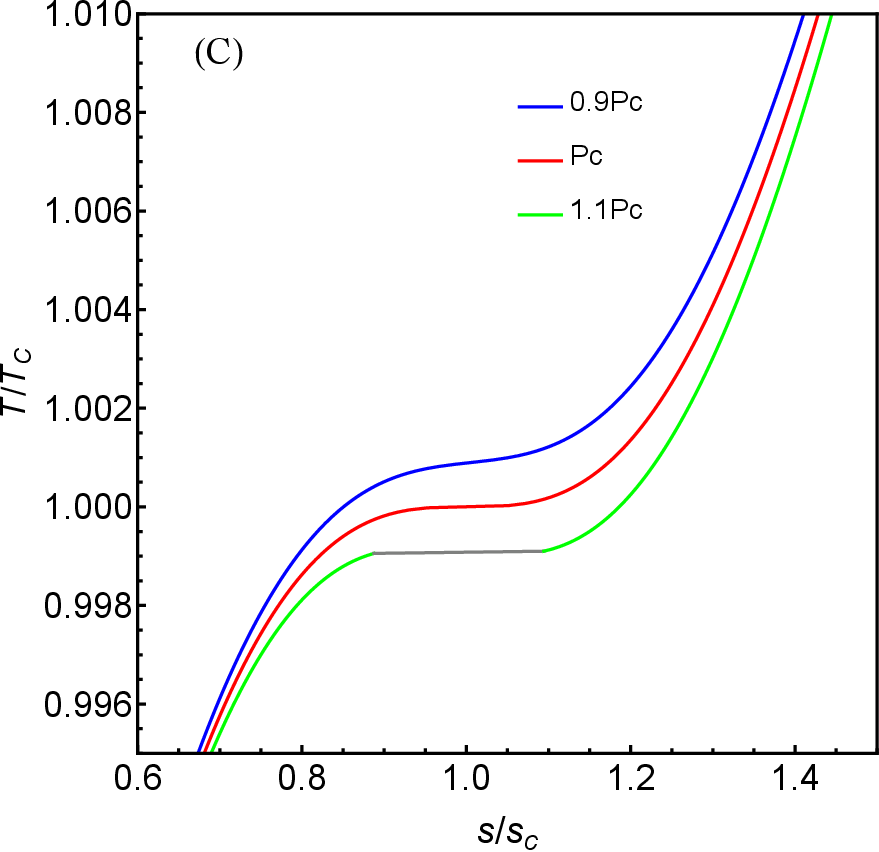}
    \includegraphics[width=0.44\columnwidth]{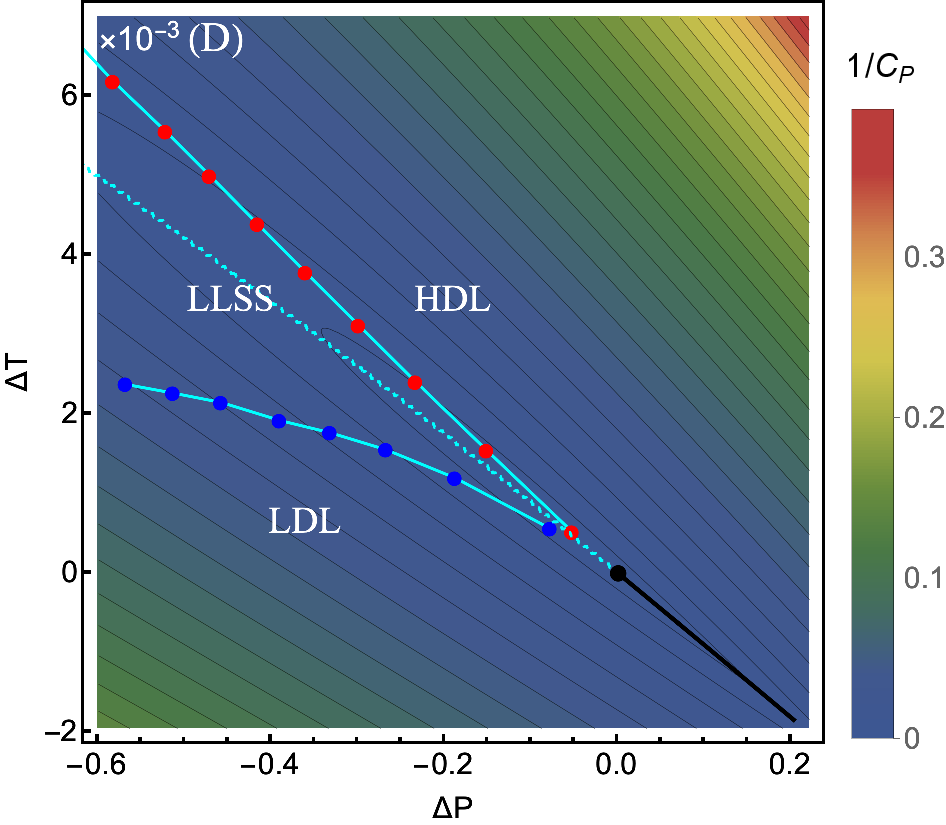}
    \caption*{Fig. S2: (A) EOSs and (B) supercritical phase diagram of the LGPT in water. Data are obtained from the NIST database.
    (C) EOSs and (D) supercritical phase diagram of the LLPT in the two-sate water model~\cite{bertrand2011peculiar, holten2012entropy} (see Sec.~S3A).}
    \label{fig:compare} 
\end{figure*}

\section*{S3. Liquid-liquid phase transitions}\label{sec:llpt}
To investigate the supercritical thermodynamics for liquid-liquid phase transitions (LLPTs), we first study a concrete two-state model of supercooled water.  Then we discuss a linear scaling theory with  a negative slope coexistence line, which can be considered as a general theoretical description of LLPTs.

\subsubsection*{A. Two-state model of supercooled water}\label{sec:twostate}

The two-state model describes the thermodynamics of a polyamorphic single component liquid, which can be regarded as a ``mixture'' of two  interconvertible states, A and B, with concentrations $1-x$ and $x$.  In water, the two states $A$ and $B$ correspond respectively to the high density liquid (HDL) and low density liquid (LDL)~\cite{poole1992phase}. We consider the mean-field version developed in Refs.~\cite{bertrand2011peculiar,holten2012entropy}.

The molar Gibbs free energy of the mixture solution reads,
\begin{equation}
\frac{G}{k_{\mathrm{B}}T}=\frac{G^{\mathrm{A}}}{k_{\mathrm{B}}T}+x\frac{G^{\mathrm{BA}}}{k_{\mathrm{B}}T}+x\ln x+(1 - x)\ln(1 - x)+\omega x(1 - x),
\label{eq:G}
\end{equation}
where $k_{\mathrm{B}}$ is the Boltzmann constant,   $x\ln x+(1 - x)\ln(1 - x)$ the mixing entropy, $\omega x(1 - x)$ the excess entropy of mixing due to interactions between A and B, and 
$\omega = 2 + \omega_0\Delta P$ with $\omega_0$ a fitting parameter. 
The term $G^{\mathrm{A}} = \sum_{m,n} c_{mn} (\Delta T)^{m} (\Delta P)^{n}$ is the Gibbs free energy of the pure state A, which can be determined by fitting experimental data with  $c_{mn}$ as adjustable coefficients. The term $G^{\mathrm{BA}} = G^{\mathrm{B}}-G^{\mathrm{A}}$  is the Gibbs free energy difference between  A and B.
Near the LLPTs, one assumes that
\begin{equation}
G^{\mathrm{BA}}/k_{\rm B}T = \lambda(\Delta T + a\Delta P + b\Delta T\Delta P)\,,
\end{equation}
where $\lambda, a, b$ are fitting parameters. At the LLPT (on the coexistence line), $G^{\mathrm{BA}} = 0$, from which one can see that 
the parameters $a$ and $b$ capture the slope and curvature of the coexistence line. The fraction $x$ is determined by the equilibrium condition

\begin{equation}
\begin{split}
\mu^{\mathrm{BA}}&=\left(\frac{\partial G}{\partial x}\right)_{T,P} \\
&=G^{\mathrm{BA}}+k_{\mathrm{B}}T\left[\ln\frac{x}{1 - x}+\omega(1 - 2x)\right] =0\,,
\end{split}
\end{equation}
whose solution is plugged into Eq.~(\ref{eq:G}). The critical values $T_{\rm c}$, $\rho_{\rm c}$  and $P_{\rm c}$ are also fitting parameters. It can be shown that this model obey critical scalings with mean-field exponents. 

With the expressions above, the EOSs can be  derived. For example, $1/\hat{\rho}(P, T) = \hat{V}(P, T) = \left(\frac{\partial \hat{G}}{\partial \hat{P}}\right)_{T}$ {and $\hat{S}(P, T) = -\left(\frac{\partial \hat{G}}{\partial \hat{T}}\right)_{P}$} using Eq.~(\ref{eq:G}). In Ref.~\cite{holten2012entropy}, the model EOSs are fitted to the published experimental data in the range of 140 K to 310 K and 0.1 MPa to 400 Mpa. The best fitting gives {$T_{\rm c} = 227.42~\rm{K}$, $P_{\rm c}=13.45~\rm{MPa}$}. Other fitting parameters can be found in~\cite{holten2012entropy}. The resulting EOSs are plotted in Fig.~S2C. 
{The $L^\pm$ lines are determined by finding the maxima of $C_P$ along paths parallel to the critical isentrope (the line of $\hat{S} = \hat{S}_{\rm c}$, where $\hat{S}_{\rm c}$ is the entropy at the critical point).}\\

\subsubsection*{B. Linear scaling theory}\label{sec:linearscailing}

\begin{figure*}[ht]
    \centering
    \includegraphics[width=0.49\columnwidth]{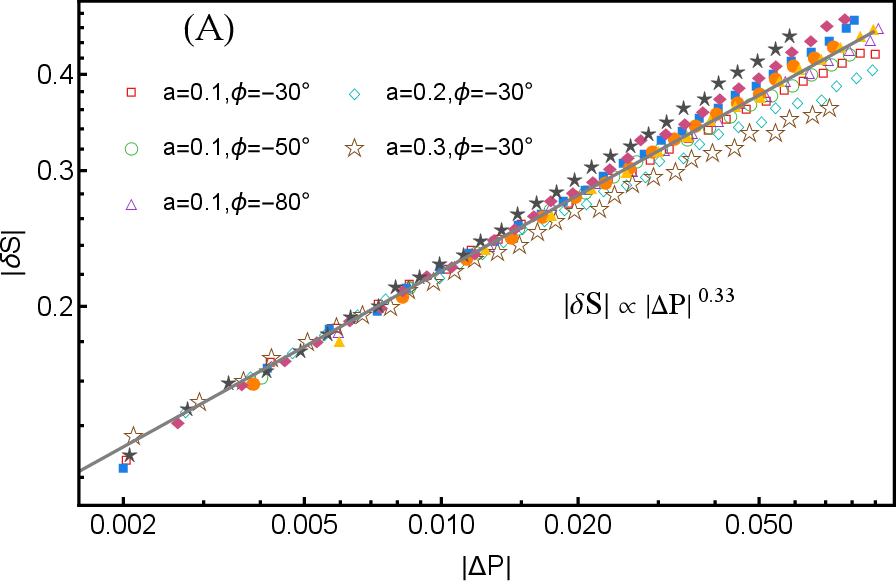}
    \includegraphics[width=0.49\columnwidth]{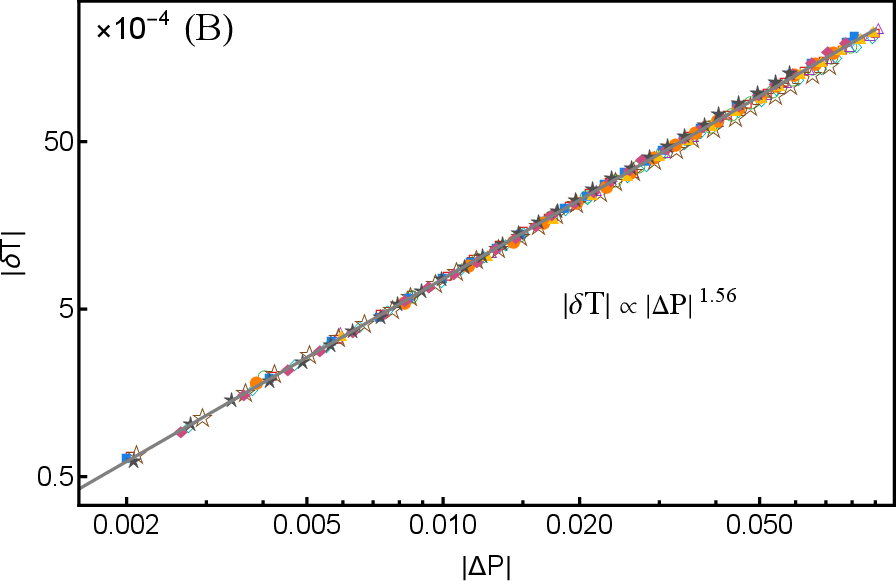}
    \caption*{Fig. S3:
    Data of $L^{\pm}$ lines given by   the linear scaling theory.
    (A) Scaling law between the reduced entropy (order parameter) $\Delta {S}$ and reduced pressure $\Delta {P}$.  (B) Scaling law between two reduced fields $\delta {T}$ and $\Delta {P}$.
 }
    \label{fig:linear_scaling} 
\end{figure*}

Near a critical point, the thermodynamic potential $\Psi(h_1, h_2)$ obeys a homogeneous scaling form
\begin{equation}
\Psi(h_1, h_2) = |h_2|^{2-\alpha} \, f\!\left( h_1/|h_2|^{\beta+\gamma} \right),
\label{eq:Psi_homogeneous}
\end{equation}
where $h_1$ and $h_2$ are the ordering and thermal scaling fields respectively, $f(x)$ is a universal scaling function, and $\alpha$, $\beta$, $\gamma$ are universal critical exponents related by the scaling law $\alpha+2\beta+\gamma=2$~\cite{Rushbrooke1963}.

Following Ref.~\cite{luo2014behavior}, the scaling fields are taken as linear combinations of the physical fileds $P$ and $T$,
\begin{equation}
\begin{aligned}
h_1 &= \Delta P\cos\varphi - \Delta T\sin\varphi,\\
h_2 &= \Delta P\sin\varphi + \Delta T\cos\varphi,
\end{aligned}
\label{eq:scaling_fields_combination}
\end{equation}
with $\Delta P = P/P_{\rm c}-1$, $\Delta T = T/T_{\rm c}-1$, and $\tan\varphi = d\hat P/d\hat T$ being the slope of the coexistence curve. For symmetric systems (\emph{e.g.} the Ising or lattice‑gas model) $\varphi=0$, so that $h_1=\Delta P$ and $h_2=\Delta T$. Real liquid–gas or liquid–liquid phase transitions are generally asymmetric, $\varphi\neq0$, and therefore the physical fields mix both $h_1$ and $h_2$.

A convenient parametric representation of the scaling fields is provided by the linear scaling theory~\cite{schofield1969parametric,schofield1969correlation}. Introducing ``polar'' variables $r$ (a measure of the distance from the critical point) and $\theta\in[-1,1]$ (an angular variable, with $\theta=\pm1$ on the coexistence curve), one writes
\begin{equation}
\begin{aligned}
h_1 &= a\, r^{\beta\delta}\,\theta\,(1-\theta^2)\,,\\[2pt]
h_2 &= r\,(1-b^2\theta^2)\,,
\end{aligned}
\label{eq:parametric_h1h2}
\end{equation}
where $\delta$ is a critical exponent linked to $\beta$ and $\gamma$ by $(\delta-1)\beta=\gamma$, $a$ is a non‑universal constant, and $b$ is a universal parameter given by $
b^2 = \frac{\delta-3}{(\delta-1)(1-2\beta)} $.
With this parametrisation, the thermodynamic potential becomes
\begin{equation}
\Psi(r,\theta) = r^{\beta(\delta+1)}\,p(\theta)\,,
\label{eq:Psi_parametric}
\end{equation}
where $p(\theta)$ is an analytic function.

The order parameter conjugate to $h_1$ is,
\begin{equation}
\phi_1 = \left(\frac{\partial\Psi}{\partial h_1}\right)_{\!h_2}
      = r^\beta\,m(\theta)\,.
\label{eq:phi1}
\end{equation}
In the linear scaling theory, one assumes $m(\theta)=k\,\theta$, with $k$ another system‑dependent parameter. The order parameter conjugate to $h_2$ is,
\begin{equation}
\phi_2 = \left(\frac{\partial\Psi}{\partial h_2}\right)_{\!h_1}
      = r^{\beta(\delta+1)-1}\,s(\theta)\,.
\label{eq:phi2}
\end{equation}
The function $s(\theta)$ is determined from the Maxwell relation
$\bigl(\partial\phi_1/\partial h_2\bigr)_{h_1}
 =\bigl(\partial\phi_2/\partial h_1\bigr)_{h_2}$,
which yields
\begin{equation}
s(\theta)=a k\bigl(s_0+s_2\theta^2\bigr)\,,
\label{eq:s_theta}
\end{equation}
with the coefficients
\begin{equation}
\begin{aligned}
s_0 &=
 -\frac{\beta\Big[-3+\delta+b^2(-1+\delta)(-2+\beta+\beta\delta)\Big]}
      {2b^4(-2+\beta+\beta\delta)(-1+\beta+\beta\delta)}\,,\\[4pt]
s_2 &=
 \frac{\beta(\delta-3)}{2b^2(-2+\beta+\beta\delta)}\,.
\end{aligned}
\label{eq:s_coefficients}
\end{equation}

The input to the theory consists of the universal critical exponents
($\beta$, $\gamma$, $\delta$, etc.) and three system‑specific parameters:
the coexistence‑line slope $\tan \varphi$, and the two constants $a$ and $k$.
Once these parameters are fixed, all thermodynamic quantities are expressed through
the two independent variables $r$ and $\theta$.  For example, the
physical fields are,
\begin{equation}
\begin{aligned}
\Delta P &=
 a r^{\beta+\gamma}\,\theta(1-\theta^2)\cos\varphi
 + r(1-b^2\theta^2)\sin\varphi\,,\\[2pt]
\Delta T &=
 r(1-b^2\theta^2)\cos\varphi
 - a r^{\beta+\gamma}\,\theta(1-\theta^2)\sin\varphi\,.
\end{aligned}
\label{eq:DeltaP_DeltaT}
\end{equation}
The reduced volume and reduced entropy are,
\begin{equation}
\Delta V
 = \left(\frac{\partial\Psi}{\partial\Delta P}\right)_{\!\Delta T}
 = -\phi_1\cos\varphi-\phi_2\sin\varphi\,,
\label{eq:DeltaV}
\end{equation}
\begin{equation}
\Delta S
 = -\left(\frac{\partial\Psi}{\partial\Delta T}\right)_{\!\Delta P}
 = -\phi_1\sin\varphi+\phi_2\cos\varphi\,.
\label{eq:DeltaS}
\end{equation}

The susceptibilities with respect to the scaling fields read
\begin{equation}
\begin{aligned}
\chi_1 &\equiv \left(\frac{\partial\phi_1}{\partial h_1}\right)_{\!h_2}
      = \frac{k}{a}\,r^{-\gamma}\,C_1(\theta)\,,\\[4pt]
\chi_2 &\equiv \left(\frac{\partial\phi_2}{\partial h_2}\right)_{\!h_1}
      = r^{-\alpha}\,C_2(\theta)\,,\\[4pt]
\chi_{12}&\equiv \left(\frac{\partial\phi_1}{\partial h_2}\right)_{\!h_1}
      = \left(\frac{\partial\phi_2}{\partial h_1}\right)_{\!h_2}
      = k\,r^{\beta-1}\,C_{12}(\theta)\,,
\end{aligned}
\label{eq:scaling_susceptibilities}
\end{equation}
where
\begin{equation}
\begin{aligned}
C_1(\theta) &=
 \frac{1-b^2\theta^2(1-2\beta)}{C_0(\theta)}\,,\\[4pt]
C_2(\theta) &=
 \frac{(1-\alpha)(1-3\theta^2)s(\theta)
       -2s_2\beta\delta\theta^2(1-\theta^2)}{C_0(\theta)}\,,\\[4pt]
C_{12}(\theta) &=
 \frac{\beta\theta\bigl[1-\delta-\theta^2(3-\delta)\bigr]}{C_0(\theta)}\,,\\[4pt]
C_0(\theta) &=
 (1-3\theta^2)(1-b^2\theta^2)
 +2\beta\delta b^2\theta^2(1-\theta^2)\,.
\end{aligned}
\label{eq:C_functions}
\end{equation}

Physical response functions are linear combinations of $\chi_1$, $\chi_2$ and
$\chi_{12}$, weighted by the mixing angle $\varphi$:
\begin{equation}
\begin{aligned}
\kappa_T &\equiv
 \left(\frac{\partial\Delta\rho}{\partial\Delta P}\right)_{\!\Delta T}
 = \chi_1\cos^2\!\varphi
   +\chi_{12}\sin2\varphi
   +\chi_2\sin^2\!\varphi\,,\\[4pt]
C_P &\equiv
 \left(\frac{\partial\Delta S}{\partial\Delta T}\right)_{\!\Delta P}
 = \chi_1\sin^2\!\varphi
   -\chi_{12}\sin2\varphi
   +\chi_2\cos^2\!\varphi\,,\\[4pt]
\alpha_P &\equiv
 \left(\frac{\partial\Delta V}{\partial\Delta T}\right)_{\!\Delta P}
 = \frac12\Big[(\chi_1-\chi_2)\sin2\varphi
                -2\chi_{12}\cos2\varphi\Big]\,.
\end{aligned}
\label{eq:physical_response}
\end{equation}

The LLPT is characterized within the linear scaling theory framework, by a negative coexistence-line slope ($\tan \varphi < 0$)~\cite{luo2014behavior}. Critical exponents are taken from the 3D Ising universality class ($\beta\simeq 0.3265$, $\gamma\simeq 1.237$)~\cite{guida1998critical}. Five sets of  parameters are explored: $\varphi= - 30^{\circ}$ with $a=k=0.1$, $0.2$, and $0.3$; $\varphi= 
- 50^{\circ}$ with $a=k=0.1$; and $\varphi= - 80^{\circ}$ with $a=k=0.1$. The $L^{\pm}$ lines computed from these sets yield the data shown in Fig.~S3. Here we also choose the entropy to be the order parameter. The $L^{\pm}$ lines are then determined from the extrema of the heat capacity along paths parallel to the critical isentrope. The results provide complementary verification of the scaling laws of $L^{\pm}$ in LLPTs, thereby strengthening the evidence for its universality and reliability.

\printbibliography
\end{document}